\documentclass{article}

\usepackage{float}

\usepackage{arxiv}
\usepackage{subcaption}
\usepackage{multirow}
\usepackage{amsmath}
\usepackage[utf8]{inputenc} 
\usepackage[T1]{fontenc}    
\usepackage{hyperref}       
\usepackage{url}            
\usepackage{booktabs}       
\usepackage{amsfonts}       
\usepackage{nicefrac}       
\usepackage{microtype}      
\usepackage{lipsum}
\usepackage{graphicx}
\graphicspath{ {./images/} }
\title{Enhanced anomaly detection in well log data through the application of ensemble GANs}

\author{
 Abdulrahman Al-Fakih* \\
  College of Petroleum Engineering and Geosciences\\
  King Fahd University of Petroleum \& Minerals\\
  Dhahran 31261, Saudi Arabia \\
  \texttt{alja2014ser@gmail.com} \\
   \And
   A. Koeshidayatullah \\
  College of Petroleum Engineering and Geosciences\\
  King Fahd University of Petroleum \& Minerals\\
  Dhahran 31261, Saudi Arabia \\
  \texttt{a.koeshidayatullah@kfupm.edu.sa} \\
  \And
 Tapan Mukerji \\
  Departments of Energy Science \& Engineering,\\ Earth \& Planetary Sciences, and Geophysics\\
  University, Stanford, CA, USA\\
  \texttt{mukerji@stanford.edu} \\
  \And
SanLinn I. Kaka \\
  College of Petroleum Engineering and Geosciences\\
  King Fahd University of Petroleum \& Minerals\\
  Dhahran 31261, Saudi Arabia \\
  \texttt{skaka@kfupm.edu.sa} \\
}

\begin{document}
\maketitle
\begin{abstract}
Although generative adversarial networks (GANs) have shown significant success in modeling data distributions for image datasets, their application to structured or tabular data, such as well logs, remains relatively underexplored. This study extends the ensemble GANs (EGANs) framework to capture the distribution of well log data and detect anomalies that fall outside of these distributions. The proposed approach compares the performance of traditional methods, such as Gaussian mixture models (GMMs), with EGANs in detecting anomalies outside the expected data distributions.  For the gamma ray (GR) dataset, EGANs achieved a precision of 0.62 and F1 score of 0.76, outperforming GMM's precision of 0.38 and F1 score of 0.54. Similarly, for travel time (DT), EGANs achieved a precision of 0.70 and F1 score of 0.79, surpassing GMM’s 0.56 and 0.71. In the neutron porosity (NPHI) dataset, EGANs recorded a precision of 0.53 and F1 score of 0.68, outshining GMM’s 0.47 and 0.61. For the bulk density (RHOB) dataset, EGANs achieved a precision of 0.52 and an F1 score of 0.67, slightly outperforming GMM, which yielded a precision of 0.50 and an F1 score of 0.65.  This work's novelty lies in applying EGANs for well log data analysis, showcasing their ability to learn data patterns and identify anomalies that deviate from them. This approach offers more reliable anomaly detection compared to traditional methods like GMM. The findings highlight the potential of EGANs in enhancing anomaly detection for well log data, delivering significant implications for optimizing drilling strategies and reservoir management through more accurate, data-driven insights into subsurface characterization. 
\end{abstract}

\section*{Keywords}
Well log data, Ensemble GANs, Anomaly detection, Gaussian mixture model, Reservoir management.

\section{Introduction}
{Well log data, such as gamma ray (GR), bulk density (RHOB), sonic travel time (DT), neutron porosity (NPHI), and deep resistivity (ILD), are fundamental to understanding subsurface geological formations \cite{darling2005well,lai2024application,luthi2001geological, ma2019petrophysical,gan2018reservoir, valentin2018estimation}. Accurate interpretation of these datasets is vital for efficient reservoir management, directly influencing drilling and production decisions \cite{mishra2022evaluation}. Proper modeling of geophysical data and the ability to detect anomalies are significant to optimizing reservoir management, with significant implications for both financial and operational outcomes. The accurate prediction of reservoir properties and anomaly detection in well log data can drive optimized drilling strategies and improve resource recovery \cite{alfakih2023reservoir, peer2017automated, struminskiy2019well}}

Anomaly detection in well log data plays a key role in identifying deviations that might indicate potential issues or opportunities for elevated resource extraction. Traditional approaches, such as Gaussian mixture models (GMMs), have been extensively employed for anomaly detection tasks in various domains due to their ability to model complex data distributions \cite{liu2019anomaly}. GMMs are probabilistic models that assume data points are generated from a mixture of Gaussian distributions. Although GMMs are effective for simple datasets, they often face challenges in dealing with the complexity and high dimensionality typical of well log data \cite{fernandes2024anomaly}. The limitations of GMMs include an inability to capture intricate data structures and a tendency to oversimplify distributions, leading to less effective anomaly detection \cite{zong2018deep}. 

Generative adversarial networks (GANs) have shown promise in the generation of synthetic data, imputing missing values, and anomaly detection in structured/ tabular data like well logs \cite{kazemi2021igani, zhang2023systematic,bourou2021review,alfakih2024a,alfakih2024b,alfakih2024c}. Common challenges such as mode collapse and limited applicability to structured data have been reported \cite{adke2022application,bhagyashree2020study}. To overcome these issues, ensemble generative adversarial networks (EGANs) was introduced, which aggregate multiple GANs to improve model robustness and performance \cite{goodfellow2014generative}. EGANs are specifically designed to mitigate mode collapse, improve stability, and capture complex data distributions more effectively than single GAN models \cite{lim2024future}.

The motivation to compare EGAN with GMM arises from the need to assess advanced anomaly detection techniques relative to traditional methods for well log data. While GMMs are well-established, they often struggle to manage the high dimensionality and complexity inherent in well log datasets, resulting in suboptimal anomaly detection. On the other hand, EGANs can offer substantial improvements by leveraging the strengths of multiple GANs to better capture complex data distributions and improve anomaly detection accuracy \cite{alfakih2024a}. This study benchmarks EGAN against GMM to quantify performance improvements in metrics such as precision, recall, and F1 score, offering a comprehensive evaluation of each model's effectiveness \cite{snorkel2022improving,ibrahim2021precision, klu2023fscore}.

A comparative summary of traditional methods, GMMs, and EGANs, including their strengths and limitations, is provided in \tableautorefname{1}. 
\begin{table}[h]
 \caption{Comparison of approaches for handling and analyzing well log data anomaly detection}
  \centering
  \begin{tabular}{|l|p{3cm}|p{3cm}|p{3cm}|l|}
    \hline
    \textbf{Method} & \textbf{Description} & \textbf{Strengths} & \textbf{Weaknesses} & \textbf{References} \\
    \hline
    GMM & Probabilistic models operate on the assumption that data points are produced from a combination of multiple Gaussian distributions & Models of multiple distributions are good for simple datasets. & Struggles with high-dimensional, complex data; can be overly simplistic. & \cite{zong2018deep} \\
    \hline
    Discriminative models & Predict missing values using conditional distributions (e.g., decision trees (DTr), Support vector machine (SVM)) & Effective for classification tasks, utilizes labeled data. & Requires labeled data, and focuses on classification rather than imputation. & \cite{marti2015anomaly,gokcesu2019sequential} \\
    \hline
    Generative models \\ (GANs \& EGANs) & Models that learn data distributions and detect anomalies by identifying deviations from predictive learned distributions. EGANs extend GANs with multiple generators for increased robustness. & Handles complex data distributions, and excels in synthetic data generation and anomaly detection. EGANs further boost robustness and capture diverse data patterns. & Computationally intensive, requires extensive training, with EGANs adding implementation complexity due to multiple models.& \cite{goodfellow2014generative, kazemi2021igani, landauer2023deep} \\
& & & & \cite{zhang2023systematic, alfakih2024a, alfakih2024b, alfakih2024c} \\
& & & & \cite{deoliveira2021synthetic, han2021gan, zhao2024novel} \\
    \hline
  \end{tabular}
  \label{tab:methods_comparison}
\end{table}

This study aims to thoroughly compare EGANs and GMMs, focusing on improvements in precision, recall, and F1 score. it also explores the potential of integrating EGANs into artificial intelligence (AI)-driven reservoir management systems for anomaly detection and more effective resource management.

The specific objectives of this research are to:

1.	Assess EGAN's performance in modeling data distributions and identifying anomalies that deviate from these distributions in well log data.

2.	Benchmark EGAN's performance against the GMM as a traditional method.

3.	Quantify the performance gains achieved by EGANs.

\section{Methodology}
\subsection{Data overview and exploratory data analysis}

The datasets used in this study comprise comprehensive well log data from two wells in the North Sea Dutch region: F17-04 and F15-A-01. These wells provide detailed logging data for GR, DT, NPHI, and RHOB logs, totaling 6553 data points within a depth range of 1925-2605 meters. These wells were selected due to their comprehensive logging data, which is pivotal for accurate machine learning model (MLM) training and evaluation.

K-Means clustering was applied jointly across the dimensions of the dataset to leverage the multivariate relationships between features. Various cluster numbers (e.g., 2, 4, 6, 8, 10, and 12) were evaluated based on visual patterns and metric consistency. Among these, 10 clusters were selected as they provided the optimal balance between capturing inherent patterns and avoiding over-segmentation or excessive generalization, as confirmed by exploratory analysis and visual assessments. The data was then standardized using StandardScaler before clustering. This standardization performed a Z-score normalization, which transformed the data such that each feature had a mean of 0 and a standard deviation of 1. This transformation ensures that the features are on the same scale, which is crucial for clustering performance. 

After applying KMeans, the dataset was filtered for clusters 0 and 1, which exhibited relevant characteristics for the anomaly detection task. Specifically, data from clusters 0 and 1 were selected, and data from other clusters were discarded. This ensures that only the data points belonging to these two clusters are retained for further analysis.

The isolation forest (IF) algorithm was employed after the clustering process to detect anomalies within the filtered data (from clusters 0 and 1). The IF algorithm flags potential anomalies by labeling data points as either -1 (anomalies) or 1 (normal). The model was trained on the training set and evaluated on the testing set, to identify outliers that can later be refined using advanced models like GMM and EGAN. 

The data was split into training and testing sets using an 80-20 split, ensuring the model was evaluated on unseen data. The results highlight the areas where anomalies occur, offering a starting point for further refinement.

The clustering results for the GR, DT, NPHI, and RHOB logs are visualized in Figure 1, where blue dots represent training samples, green dots indicate normal testing samples and red stars highlight anomalies detected by the IF algorithm. These visualizations provide insight into the clustering process and the detection of potential anomalies, illustrating the data distribution, clustering effectiveness, and the initial anomaly detection results.

\begin{figure}[htbp]
  \centering
  \begin{minipage}{0.45\textwidth}
    \centering
    \includegraphics[width=\linewidth]{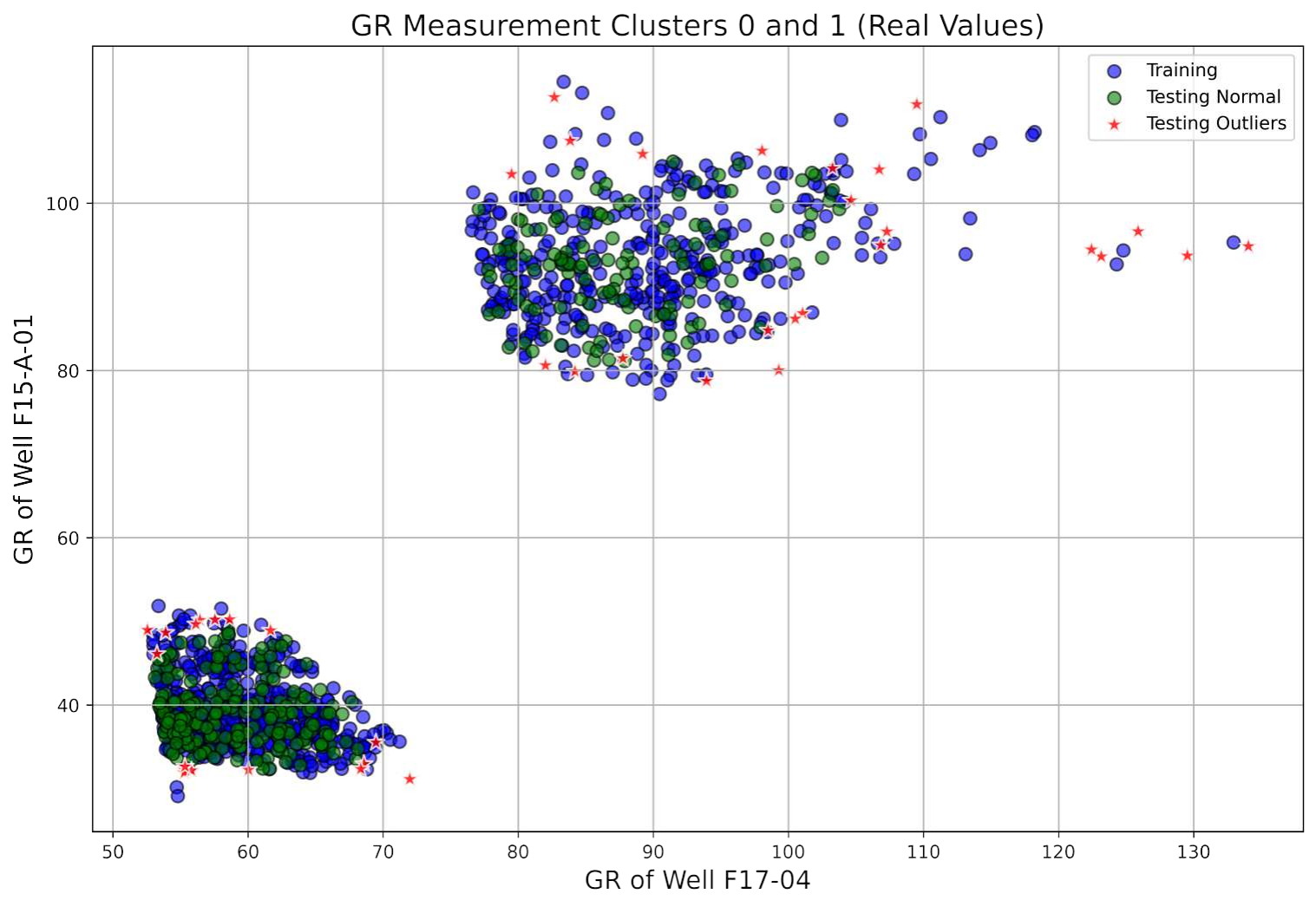} 
    \subcaption{ GR dataset} \label{fig:gr}
  \end{minipage}
  \hspace{0.5cm}
  \begin{minipage}{0.45\textwidth}
    \centering
    \includegraphics[width=\linewidth]{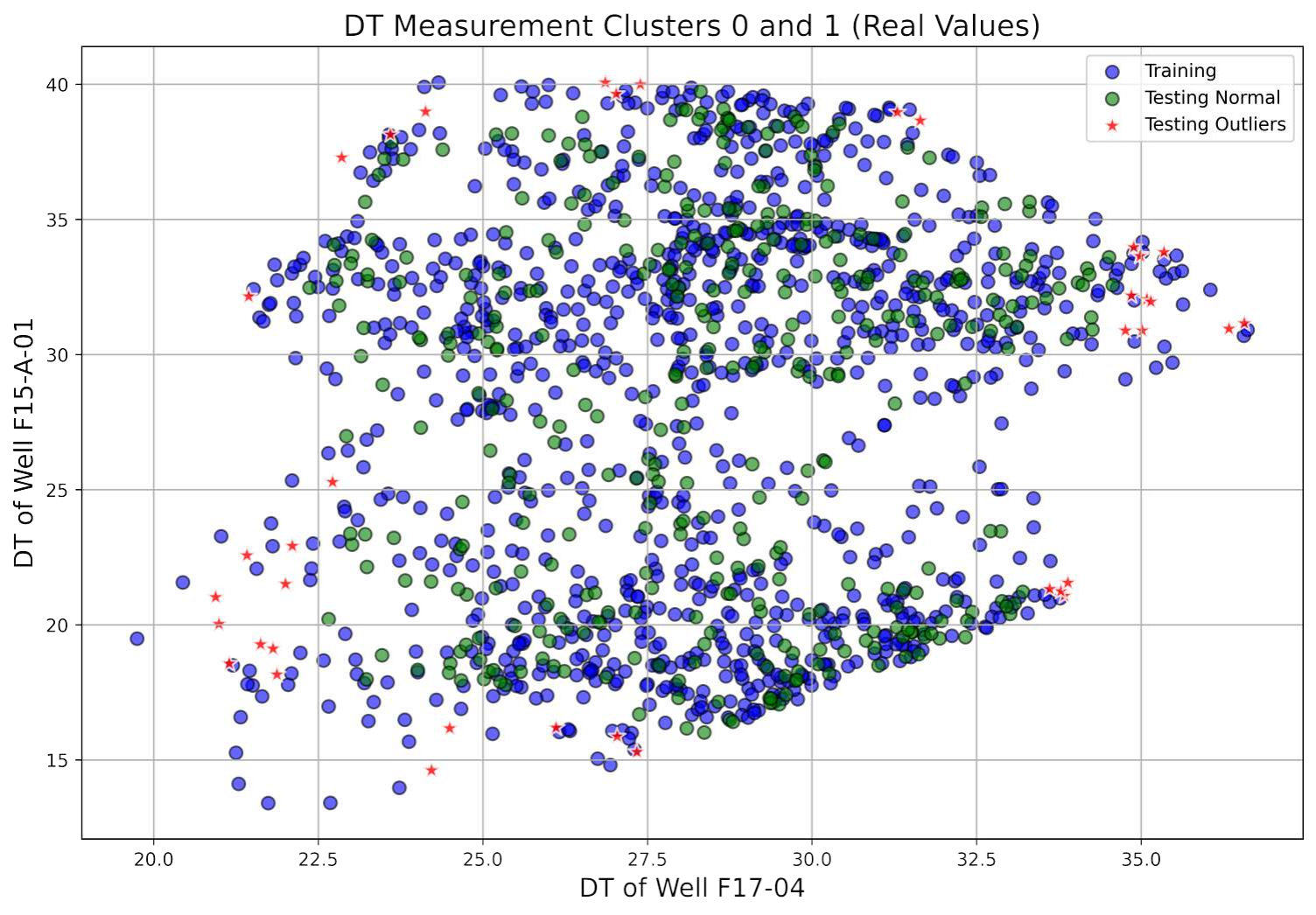} 
    \subcaption{DT dataset} \label{fig:dt}
  \end{minipage}

  \vspace{1cm}

  \begin{minipage}{0.45\textwidth}
    \centering
    \includegraphics[width=\linewidth]{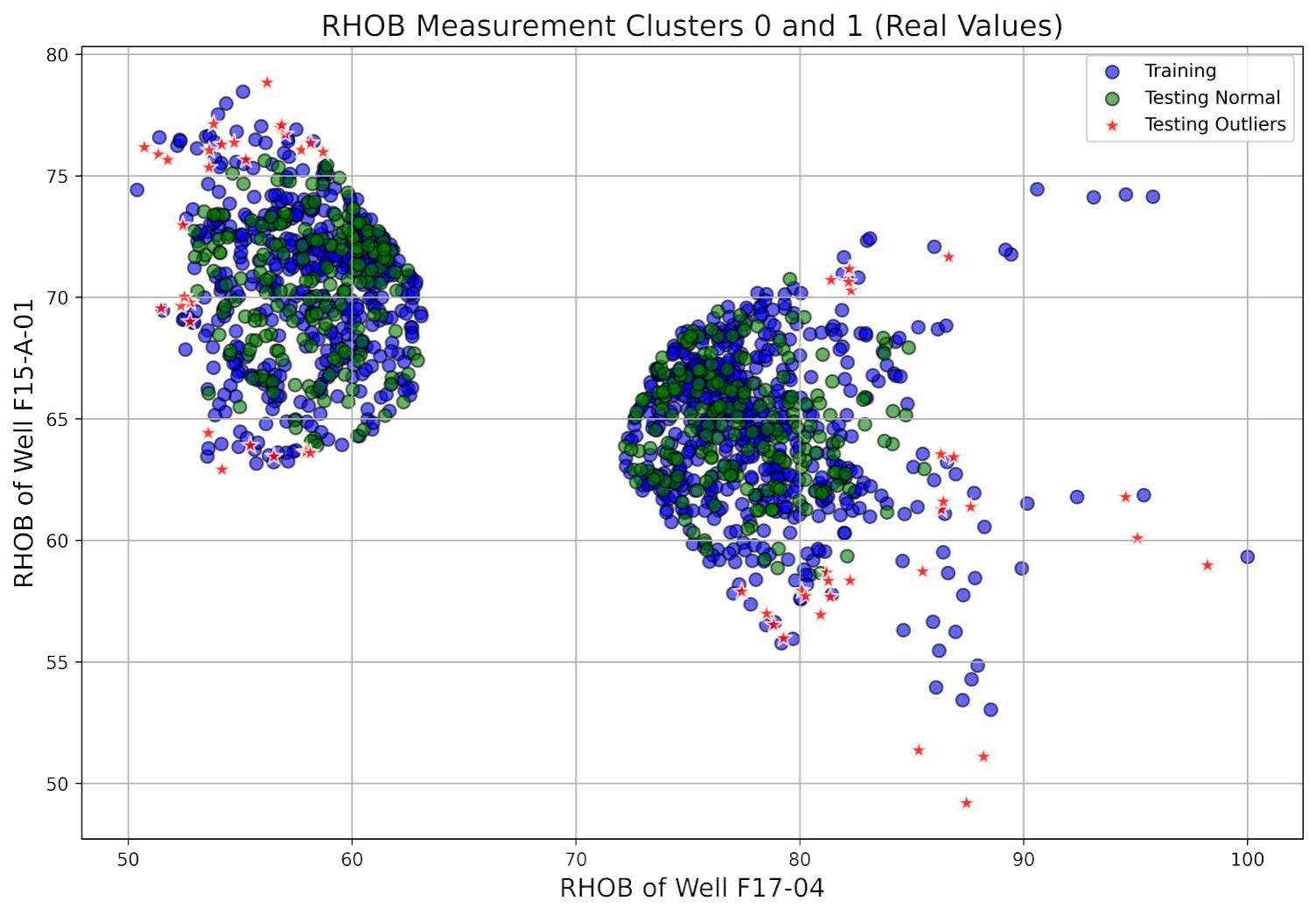} 
    \subcaption{ RHOB dataset} \label{fig:rhob}
  \end{minipage}
  \hspace{0.5cm}
  \begin{minipage}{0.45\textwidth}
    \centering
    \includegraphics[width=\linewidth]{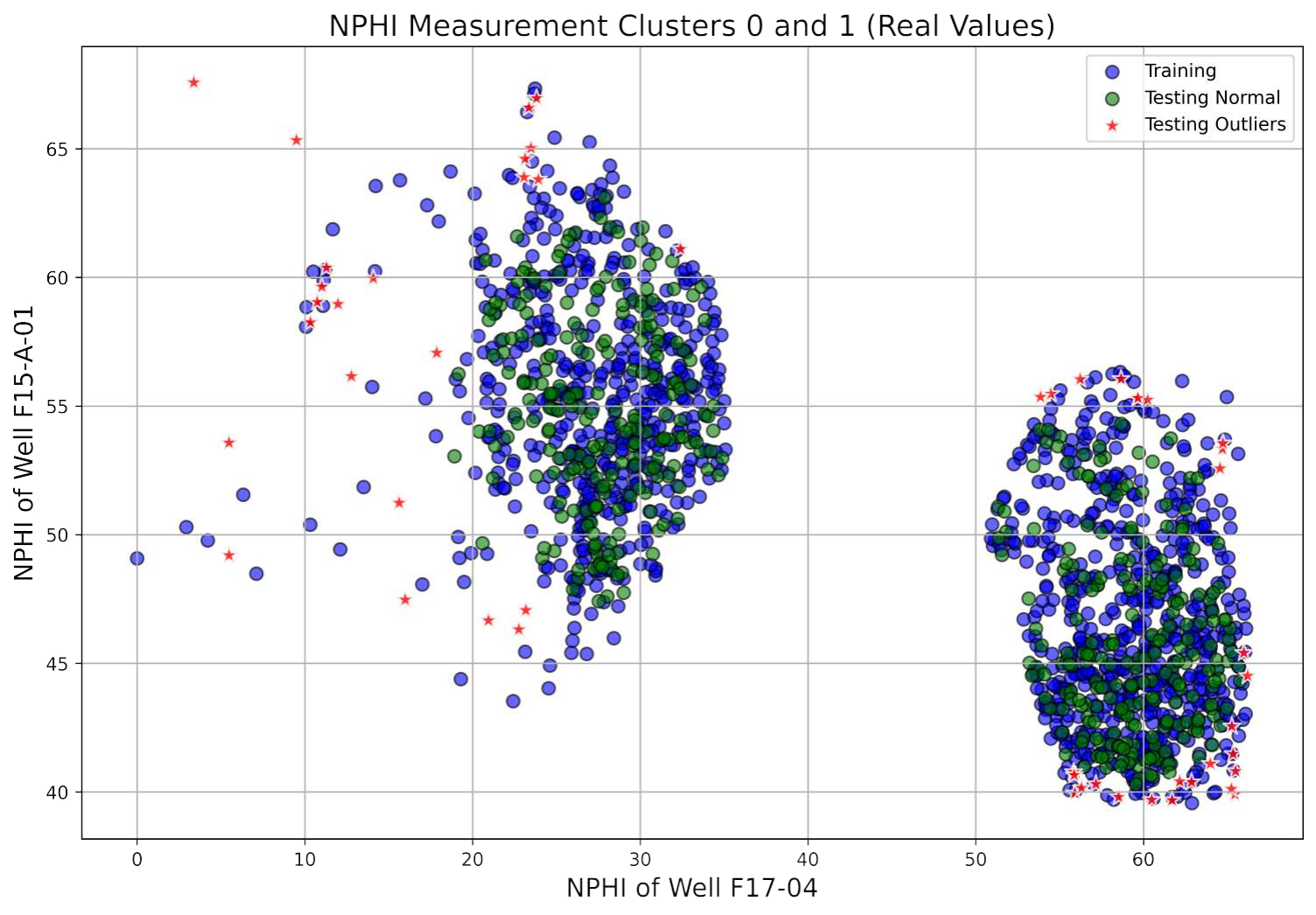} 
    \subcaption{ NPHI dataset} \label{fig:nphi}
  \end{minipage}
  
  \caption{Visualization of dataset distributions and clustering results for (a) GR, (b) DT, (c) RHOB, and (d) NPHI logs. The blue dots represent training samples, the green dots indicate normal testing samples, and the red circles denote anomaly samples detected by the IF algorithm.}
  \label{fig:datasets}
\end{figure}

\subsection{Model configuration and hyperparameter settings}
Each anomaly detection model GMM and EGAN was configured with tailored hyperparameters to optimize performance. The chosen hyperparameters are important as they directly impact the models' accuracy, efficiency, and effectiveness in detecting anomalies. \tableautorefname{2} outlines the specific hyperparameter settings for each model, providing transparency and facilitating reproducibility of the study's methodology.

\begin{table}[h]
 \caption{Hyperparameter settings for anomaly detection models}
  \centering
  \begin{tabular}{|l|l|p{7cm}|l|}
    \hline
    \textbf{Method} & \textbf{Hyperparameter} & \textbf{Description} & \textbf{Value} \\
    \hline
    \multirow{8}{*}{GMM} & Cluster number & Number of Gaussian components & 2 \\
    \cline{2-4}
    & Covariance type & Type of covariance parameters & Full \\
    \cline{2-4}
    & Max iterations & Maximum number of iterations for the expectation-maximization (EM) algorithm & 100 \\
    \cline{2-4}
    & Tol & Convergence threshold & 1e-3 \\
    \cline{2-4}
    & Init params & Method to initialize the weights & K-means \\
    \cline{2-4}
    & Random state & Seed for random number generation & 42 \\
    \cline{2-4}
    & Evaluation metric & Metrics used for evaluation & Precision, recall, F1 \\
    \cline{2-4}
    & Visualization & Methods for visualizing results & Contour plots, GMM results \\
    \hline
    \multirow{12}{*}{EGANs} & Generator hidden layers & The count of hidden layers within the generator. & 2 \\
    \cline{2-4}
    & G neurons & The number of neurons in each hidden layer of the generator. & 128 \\
    \cline{2-4}
    & D hidden layers & The number of hidden layers in the discriminator. & 2 \\
    \cline{2-4}
    & D neurons & The number of neurons in the discriminator. & 128 \\
    \cline{2-4}
    & Learning rate (G) & The learning rate for the generator & 0.001 - 0.002 \\
    \cline{2-4}
    & Learning rate (D) & The learning rate for discriminator & 0.001 - 0.002 \\
    \cline{2-4}
    & Batch Size & Number of samples processed per iteration & 32 - 64 \\
    \cline{2-4}
    & Number of training steps & Total iterations for training & 10,000 \\
    \cline{2-4}
    & Critic Iterations & Number of discriminator updates per generator update & 3 \\
    \cline{2-4}
    & Loss function & Loss function used & Binary cross-entropy \\
    \cline{2-4}
    & Evaluation metric & Metrics used for evaluation & Precision, recall, F1 \\
    \cline{2-4}
    & Visualization & Methods for visualizing results & Contour plots, real vs. fake data \\
    \hline
  \end{tabular}
  \label{tab:hyperparameters}
\end{table}

\subsection{Proposed models’ workflow}
The overall workflow for this study involved two key models: GMM and EGANs. Figure 2 illustrates the distinct workflows used for each model. Figure 2(a) represents the anomaly detection workflow for GMM, showcasing steps such as clustering, model training, evaluation, and result visualization. Figure 2(b) presents the workflow for EGANs, highlighting the processes of setting hyperparameters, defining network architectures, training the EGAN model, and evaluating its performance.

\begin{figure}[htbp] 
    \centering
    \includegraphics[width=2.2\textwidth]{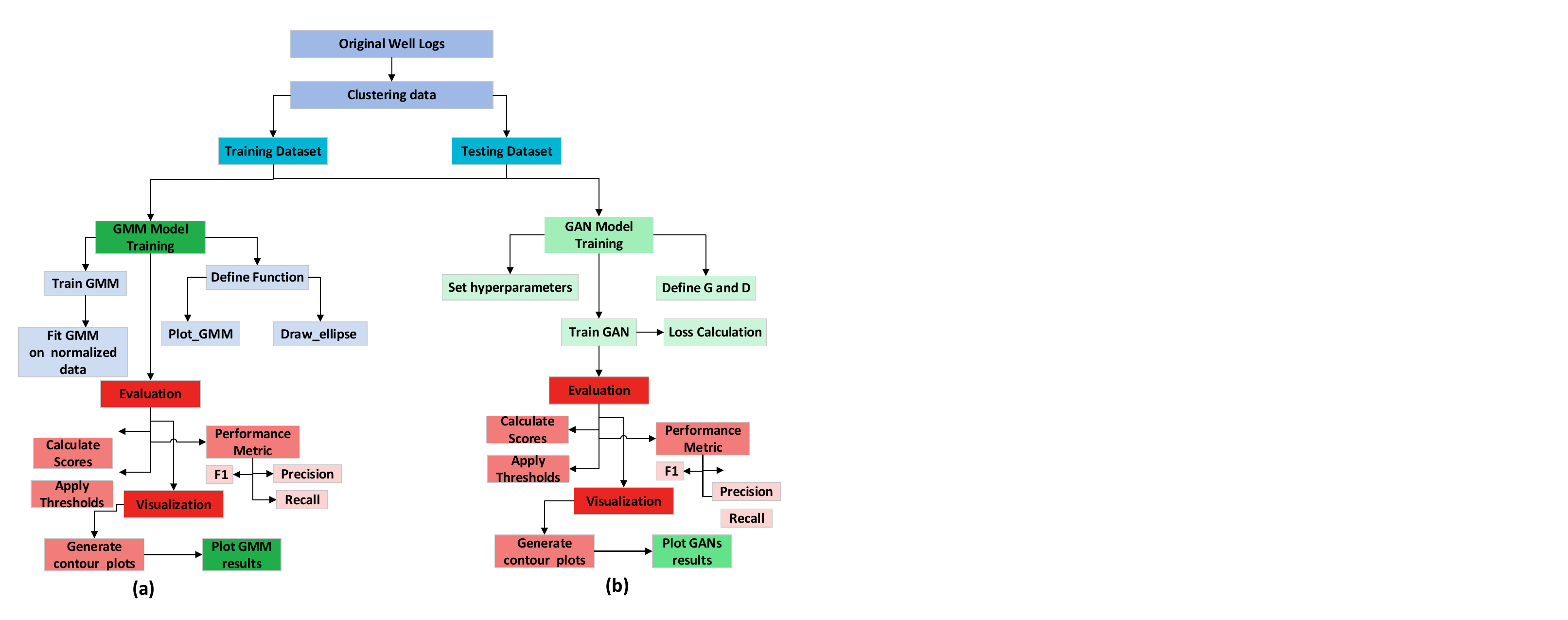}  
    \caption{Workflow for anomaly detection in well log data using (a) Gaussian Mixture Model (GMM) and (b) Ensemble Generative Adversarial Networks (EGANs).}
    \label{fig:fig1}
\end{figure}

\subsection{Technical description for each model}
\subsubsection{Gaussian mixture model}
The GMM is a probabilistic model that posits data points originate from a blend of multiple Gaussian distributions, each characterized by unknown parameters\cite{stauffer1999adaptive}. Given a d-dimensional vector x, the probability density function of the Gaussian mixture model can be formulated as in equation 1:

\begin{equation}
p(x) = \sum_{i=1}^{c} w_i p_i(x)
\end{equation}

In this formulation, \( c \) denotes the number of mixture components, and the mixture weights \( w_i \) adhere to the condition
\[
\sum_{i=1}^c w_i = 1 \quad \text{and} \quad w_i > 0 \, \text{for each component} \, i.
\]
Each component density \( p_i(x) \) represents the probability density function of a Gaussian distribution characterized by a \( d \times 1 \) mean vector \( \mu_i \) and a \( d \times d \) covariance matrix \( \Sigma_i \) as shown in equation 2:
\begin{equation}
p(x) = \frac{1}{(2\pi)^{d/2} |\Sigma_i|^{1/2}} \exp \left\{-\frac{1}{2} (x - \mu_i)^T \Sigma_i^{-1} (x - \mu_i) \right\} 
\end{equation}

GMMs are capable of approximating any continuous probability density function with high precision, provided that an adequate number of Gaussian components are used, and their means, covariances, and weights are properly adjusted \cite{reynolds2015gaussian}. This makes GMMs suitable for modeling complex data distributions, especially in lower-dimensional settings. However, when applied to high-dimensional data or data with intricate structures, GMMs may struggle to capture the full complexity of the distribution, potentially oversimplifying it. In such cases, alternative methods may be required to better model the underlying data patterns \cite{azizyan2013minimax}.

\subsubsection{Generative adversarial networks model}
\lipsum[6]

GANs, first proposed by Goodfellow et al. (2014), consist of two neural networks in competition: the generator (G) and the discriminator (D). Through an adversarial training process, the generator aims to produce synthetic data that resembles real data, while the discriminator aims to differentiate between real and synthetic data. This adversarial training framework progressively upgrades the generator's capacity to generate realistic data samples \cite{goodfellow2014generative}.

The GAN architecture involves two main components, the G and the D. Both are implemented as three-layer multi-layer perceptrons (MLPs), each with distinct roles and configurations. The G generates synthetic data samples, and the D evaluates the authenticity of these samples by distinguishing real data from the synthetic data generated by G.

The adversarial training process in GANs is formulated as a minimax game, where the generator and discriminator play against each other. The value function \( V(G,D) \) governing this game is defined as follows in equation 3:

\begin{equation}
\min_G \max_D V(G,D) = \mathbb{E}_{x \sim P_{\text{data}}} [ \log D(x)] + \mathbb{E}_{z \sim P_z(z)} [ \log (1 - D(G(z)))]
\end{equation}

Where \( D(x) \) represents the discriminator's estimation of the probability that the input \( x \) is real data, \( G(z) \) is the generator's output given noise \( z \), \( P_{\text{data}} \) is the data distribution, and \( P_z \) is the noise distribution.

Here, \( G \) strives to maximize the probability that \( D \) misclassifies synthetic data as real, while \( D \) aims to minimize its classification error. This adversarial training continues iteratively, improving the generator's ability to generate data that is indistinguishable from real data.

For anomaly detection, after the GAN is trained, it is the discriminator that is typically used to detect anomalies. In the context of anomaly detection, outliers or anomalies are considered as synthetic data that \( D \) classifies as fake (or with a low probability of being real). Thus, the discriminator is used to evaluate whether new data points are outliers by assessing whether they fit the learned distribution of real data.

\subsection{Evaluation metrics for each model configuration}

To comprehensively evaluate the effectiveness of the models employed in this study, several key metrics were employed. These metrics offer insights into various facets of the model's predictive accuracy and reliability, particularly in the context of anomaly detection in well log data. The following metrics were used:

\begin{itemize}
    \item \textbf{Precision (Prec)}: This metric measures the ratio of true positive predictions to the total predicted positives. It is key for understanding the proportion of relevant instances among the retrieved instances as shown in equation (4) \cite{rebala2019introduction} \cite{powers2007evaluation}.
    \begin{equation}
    \text{Precision} = \frac{\text{True Positives (TP)}}{\text{True Positives (TP)} + \text{False Positives (FP)}} 
    \end{equation}

    \item \textbf{Recall (Rec)}: Also known as sensitivity, this metric measures the ratio of true positive predictions to the total number of actual positives. It helps in understanding the proportion of actual positives that the model correctly identified, as shown in equation (5).
    \begin{equation}
    \text{Recall} = \frac{\text{True Positives (TP)}}{\text{True Positives (TP)} + \text{False Negatives (FN)}} 
    \end{equation}
    
    \item \textbf{F1 Score (F1)}: The F1 Score, derived as the harmonic mean of precision and recall, serves as a unified measure that addresses both false positives and false negatives. Its utility is particularly evident in scenarios with imbalanced class distributions, as illustrated in equation (6).
    \begin{equation}
    F1 = 2 \times \frac{\text{Precision} \times \text{Recall}}{\text{Precision} + \text{Recall}} 
    \end{equation}

    \item \textbf{Accuracy (Acc)}: Accuracy quantifies the proportion of correctly predicted instances (both true positives and true negatives) out of all instances. It provides an overall indication of the model's correctness, as depicted in equation (7).
    \begin{equation}
    \text{Accuracy} = \frac{\text{True Positives (TP)} + \text{True Negatives (TN)}}{\text{Total Instances}} 
    \end{equation}
\end{itemize}

These evaluation metrics are essential for a comprehensive assessment of MLM, particularly in scenarios involving anomaly detection and classification. They help understand the models' strengths and weaknesses, guiding improvements and optimizations in future work. Table 3 summarizes the detailed descriptions and references for these metrics.

\begin{table}[h]
  \caption{Summary of evaluation metrics for model performance}
  \centering
  \begin{tabular}{|l|l|p{7cm}|l|}
    \hline
    \textbf{Metric} & \textbf{Abbreviation} & \textbf{Description} & \textbf{References} \\
    \hline
    Precision & Prec & Accuracy of positive predictions & \cite{rebala2019introduction,powers2007evaluation} \\
    \hline
    Recall & Rec & Ability to identify all relevant instances & \cite{rebala2019introduction,powers2007evaluation}\\
    \hline
    F1 Score & F1 & The harmonic mean of precision and recall. & \cite{rebala2019introduction} \\
    \hline
    Accuracy & Acc & Overall correctness of the model & \cite{powers2007evaluation} \\
    \hline
  \end{tabular}
  \label{tab:evaluation_metrics}
\end{table}

\section{Results and discussion}
The comparative study evaluated two advanced MLMs, EGANs, and GMM, using well log data from two North Sea Dutch wells, focusing on GR, DT, RHOB, and NPHI logs. The objective was to assess each model’s ability to classify and detect anomalies in these datasets.
\subsection{Gaussian mixture model analysis}
The GMM analysis for the GR, DT, RHOB, and NPHI logs reveals distinct patterns and insights, as shown in Figure 3 with its four-column layout.
\begin{figure}[htbp]
\centering
\begin{tabular}{|c|c|c|c|}
\hline
\textbf{Datasets} & \textbf{Training Stage} & \textbf{Data Distribution} & \textbf{Testing Stage} \\
\hline
\begin{minipage}{0.24\textwidth}
    \centering
    \includegraphics[width=\textwidth]{pd_Figures/Figure1_a.pdf}
    {\textbf{GR}}
\end{minipage} &
\begin{minipage}{0.24\textwidth}
    \centering
    \includegraphics[width=\textwidth]{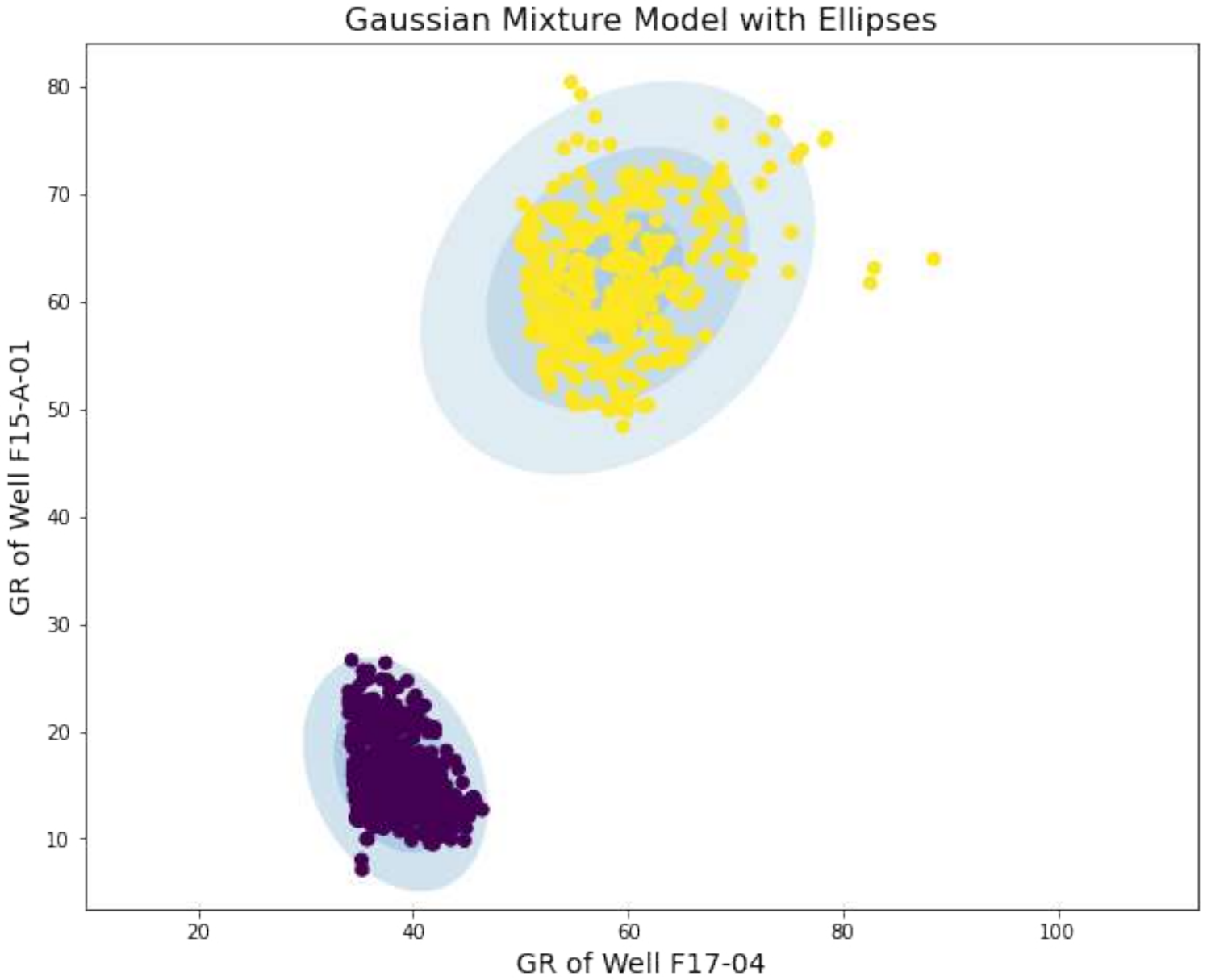}
\end{minipage} &
\begin{minipage}{0.26\textwidth}
    \centering
    \includegraphics[width=\textwidth]{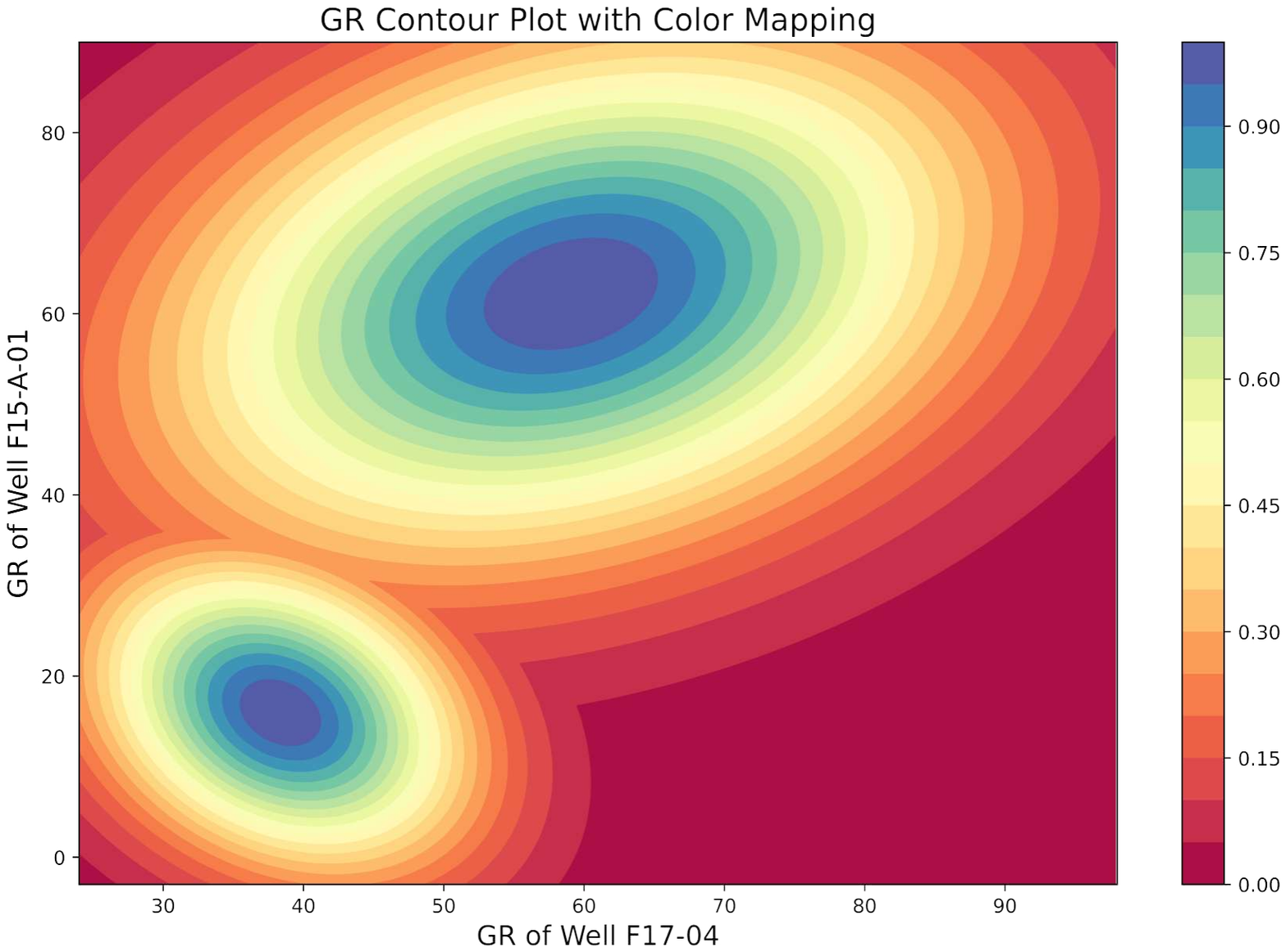}
\end{minipage} &
\begin{minipage}{0.26\textwidth}
    \centering
    \includegraphics[width=\textwidth]{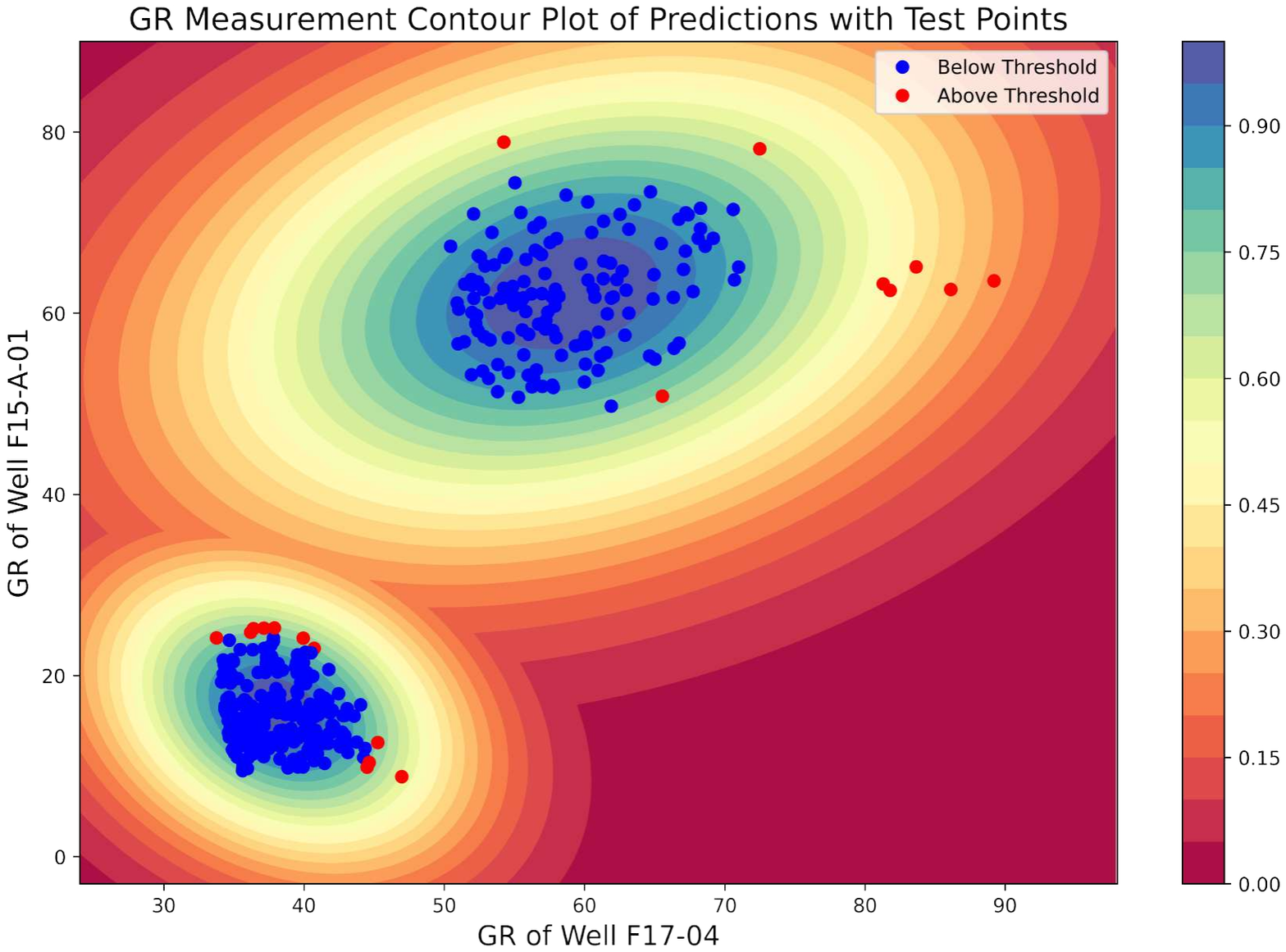}
\end{minipage} \\
\hline
\begin{minipage}{0.24\textwidth}
    \centering
    \includegraphics[width=\textwidth]{pd_Figures/Figure1_b.pdf}
    {\textbf{DT}}
\end{minipage} &
\begin{minipage}{0.24\textwidth}
    \centering
    \includegraphics[width=\textwidth]{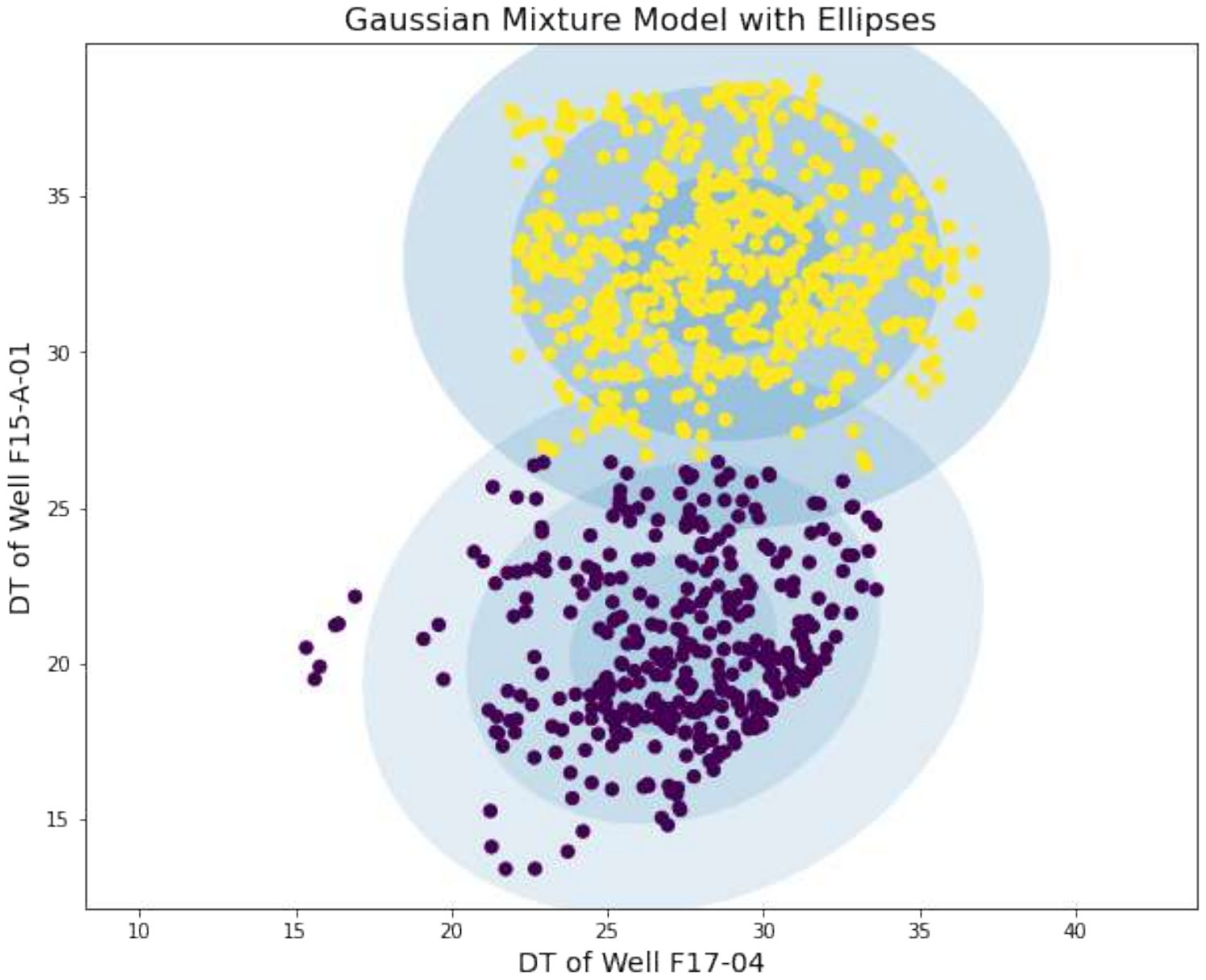}
\end{minipage} &
\begin{minipage}{0.26\textwidth}
    \centering
    \includegraphics[width=\textwidth]{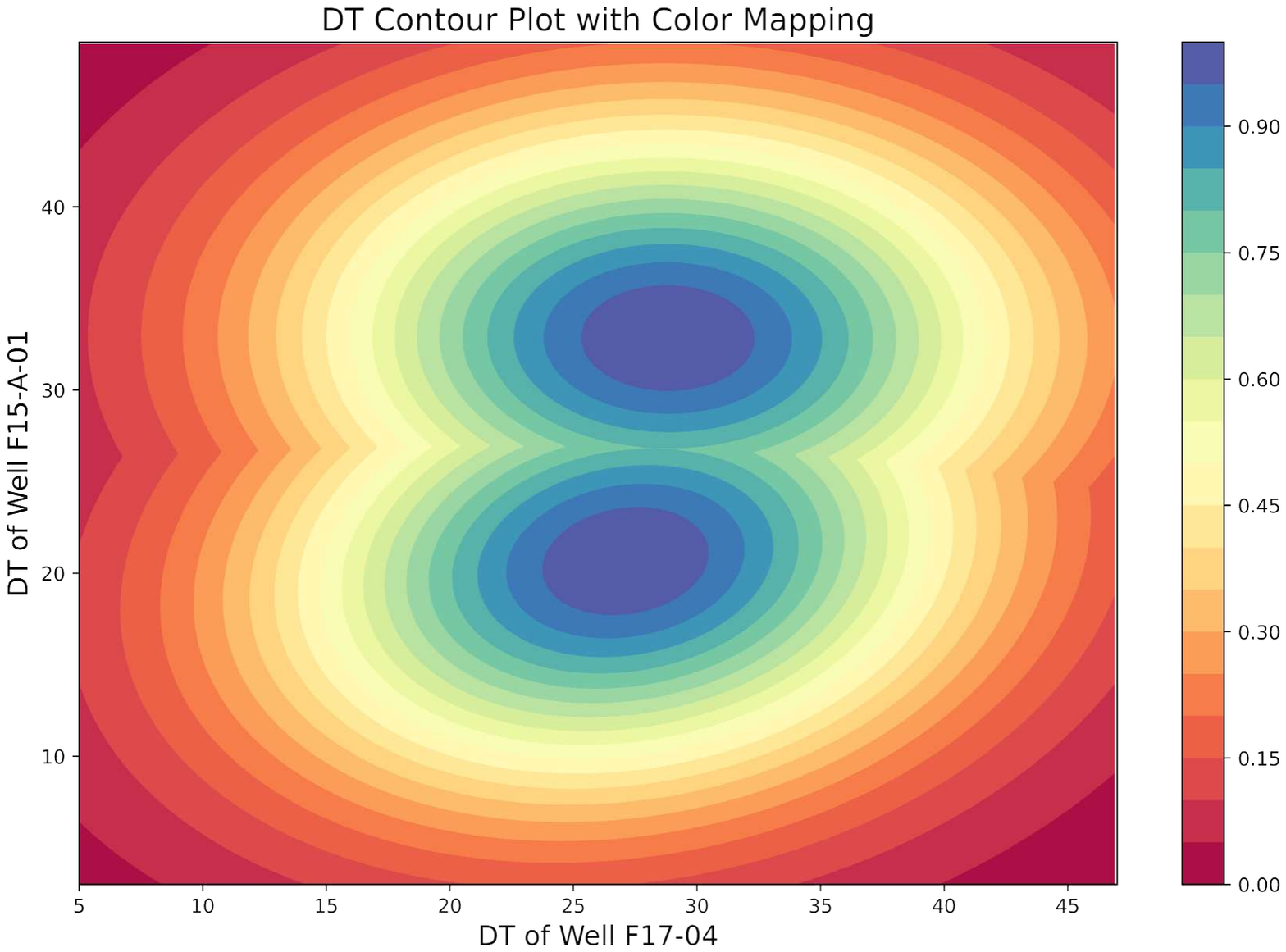}
\end{minipage} &
\begin{minipage}{0.26\textwidth}
    \centering
    \includegraphics[width=\textwidth]{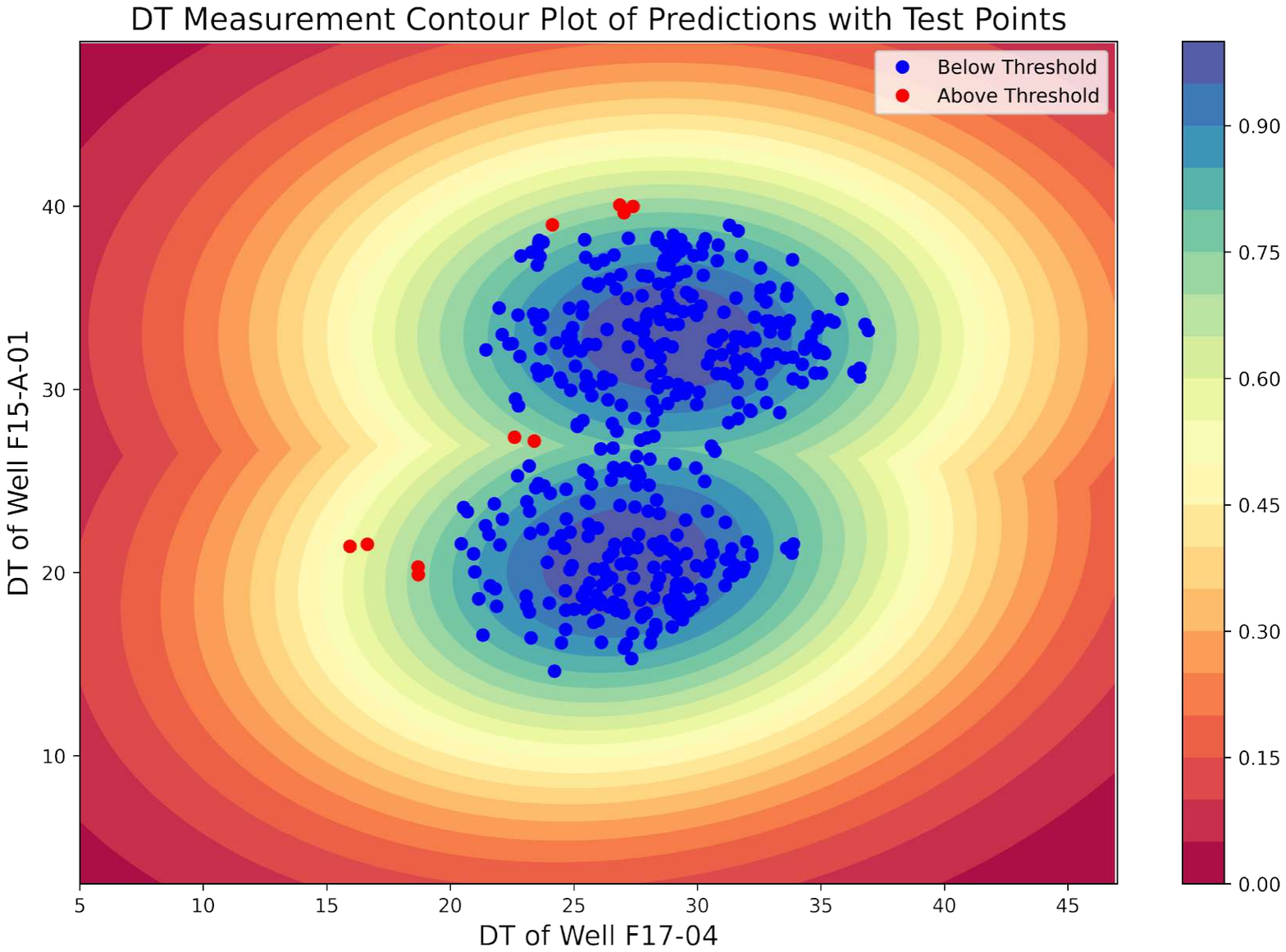}
\end{minipage} \\
\hline
\begin{minipage}{0.24\textwidth}
    \centering
    \includegraphics[width=\textwidth]{pd_Figures/Figure1_c.pdf}
    {\textbf{RHOB}}
\end{minipage} &
\begin{minipage}{0.24\textwidth}
    \centering
    \includegraphics[width=\textwidth]{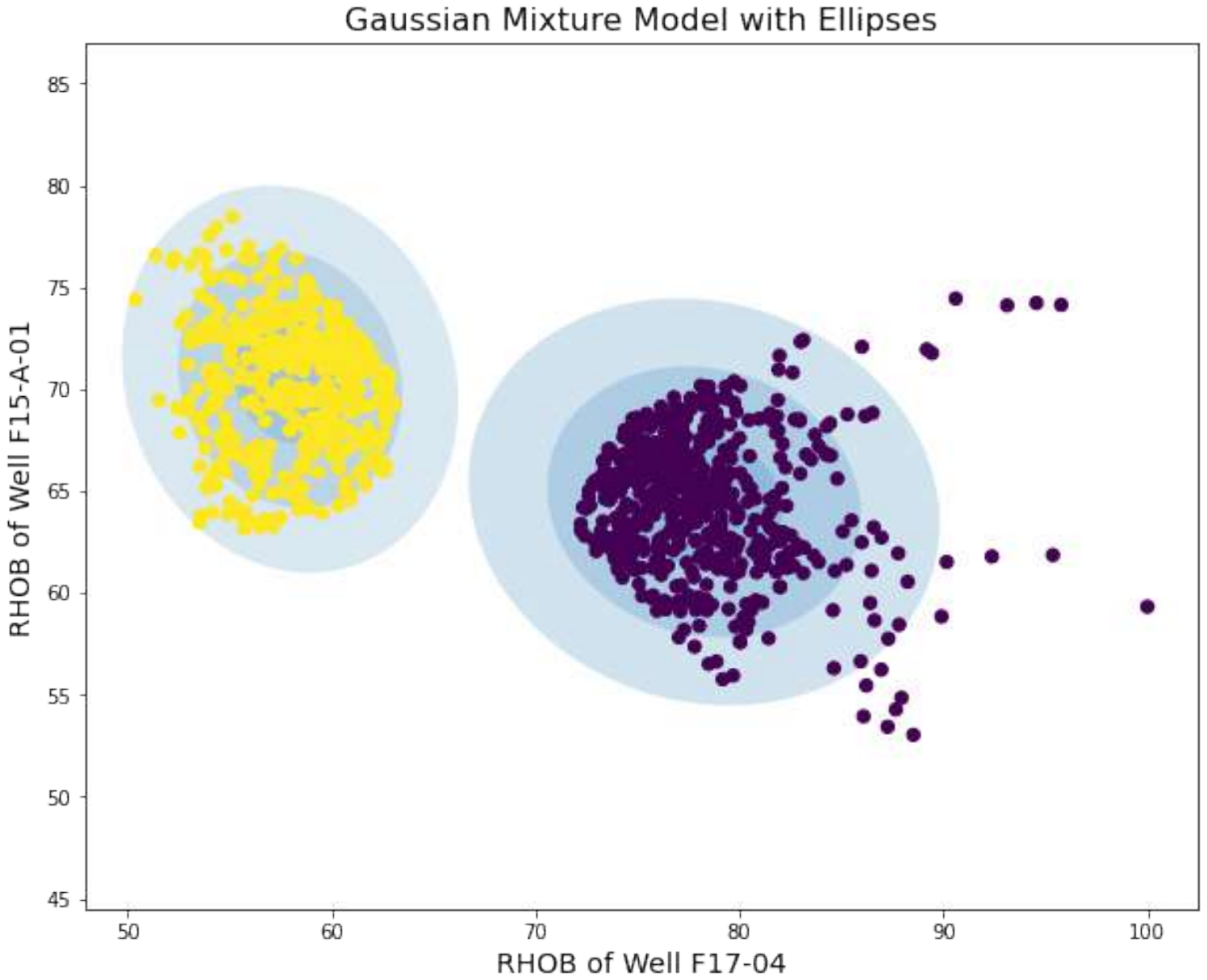}
\end{minipage} &
\begin{minipage}{0.26\textwidth}
    \centering
    \includegraphics[width=\textwidth]{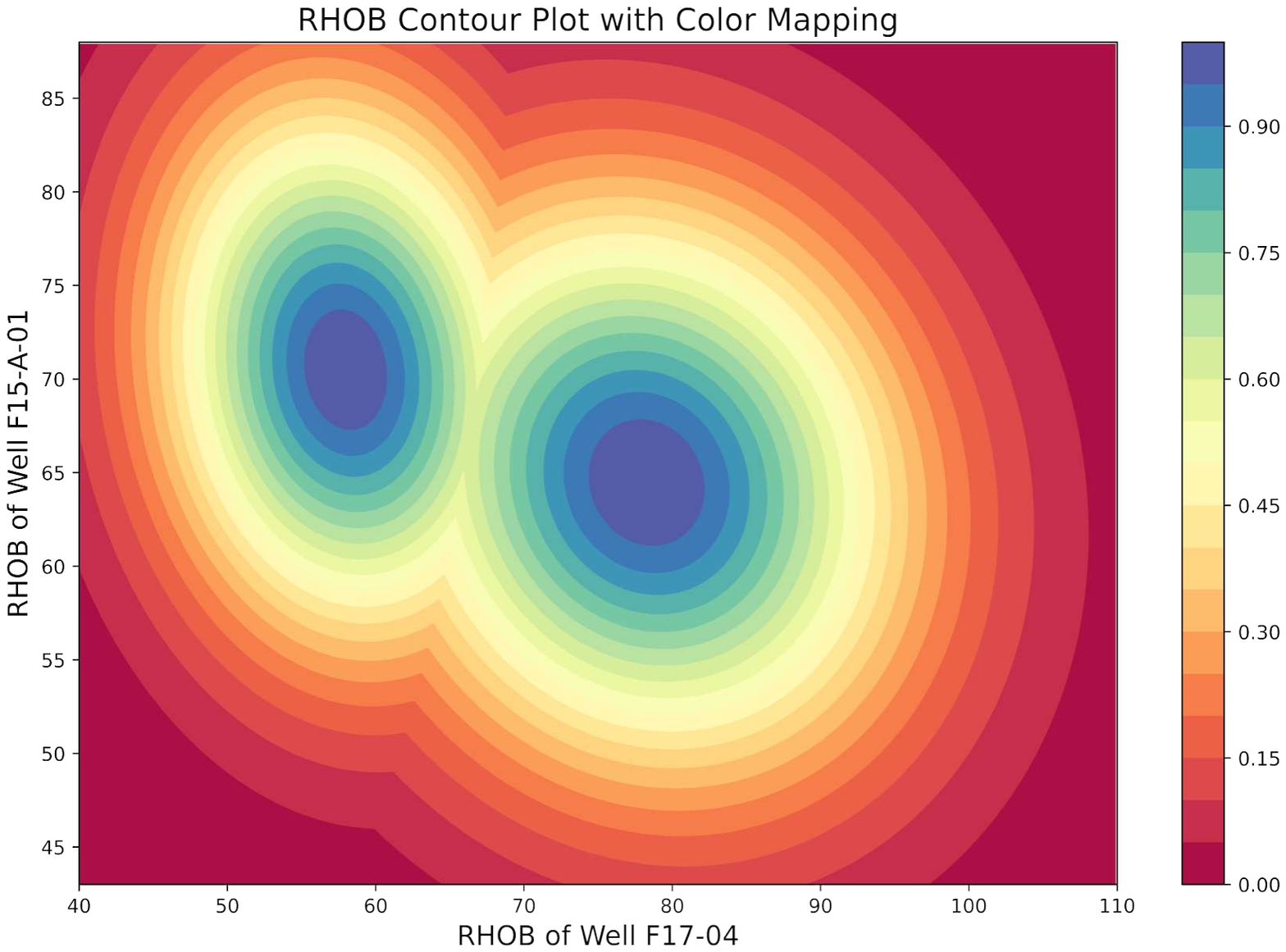}
\end{minipage} &
\begin{minipage}{0.26\textwidth}
    \centering
    \includegraphics[width=\textwidth]{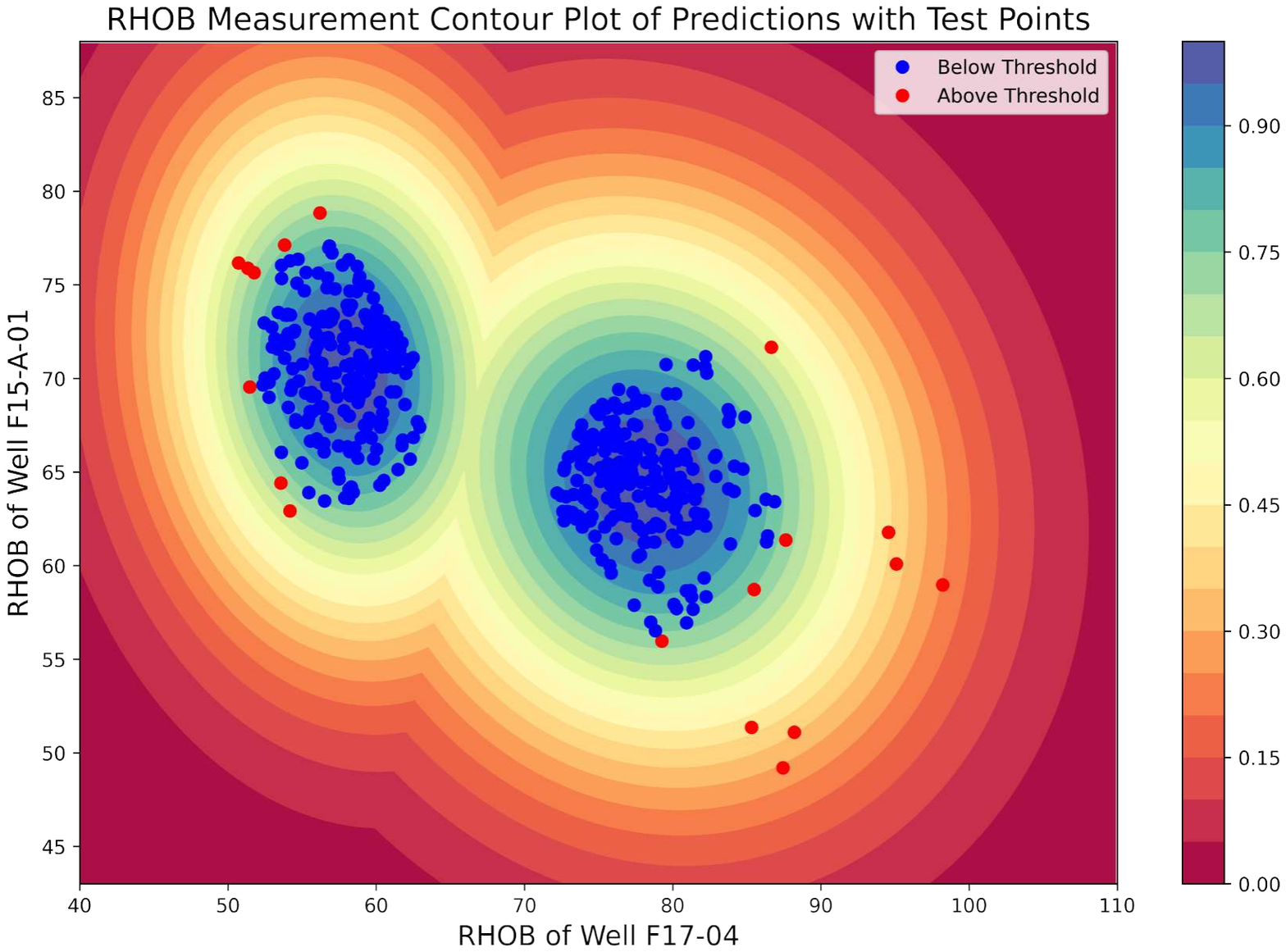}
\end{minipage} \\
\hline
\begin{minipage}{0.24\textwidth}
    \centering
    \includegraphics[width=\textwidth]{pd_Figures/Figure1_d.pdf}
{\textbf{NPHI}}
\end{minipage} &
\begin{minipage}{0.24\textwidth}
    \centering
    \includegraphics[width=\textwidth]{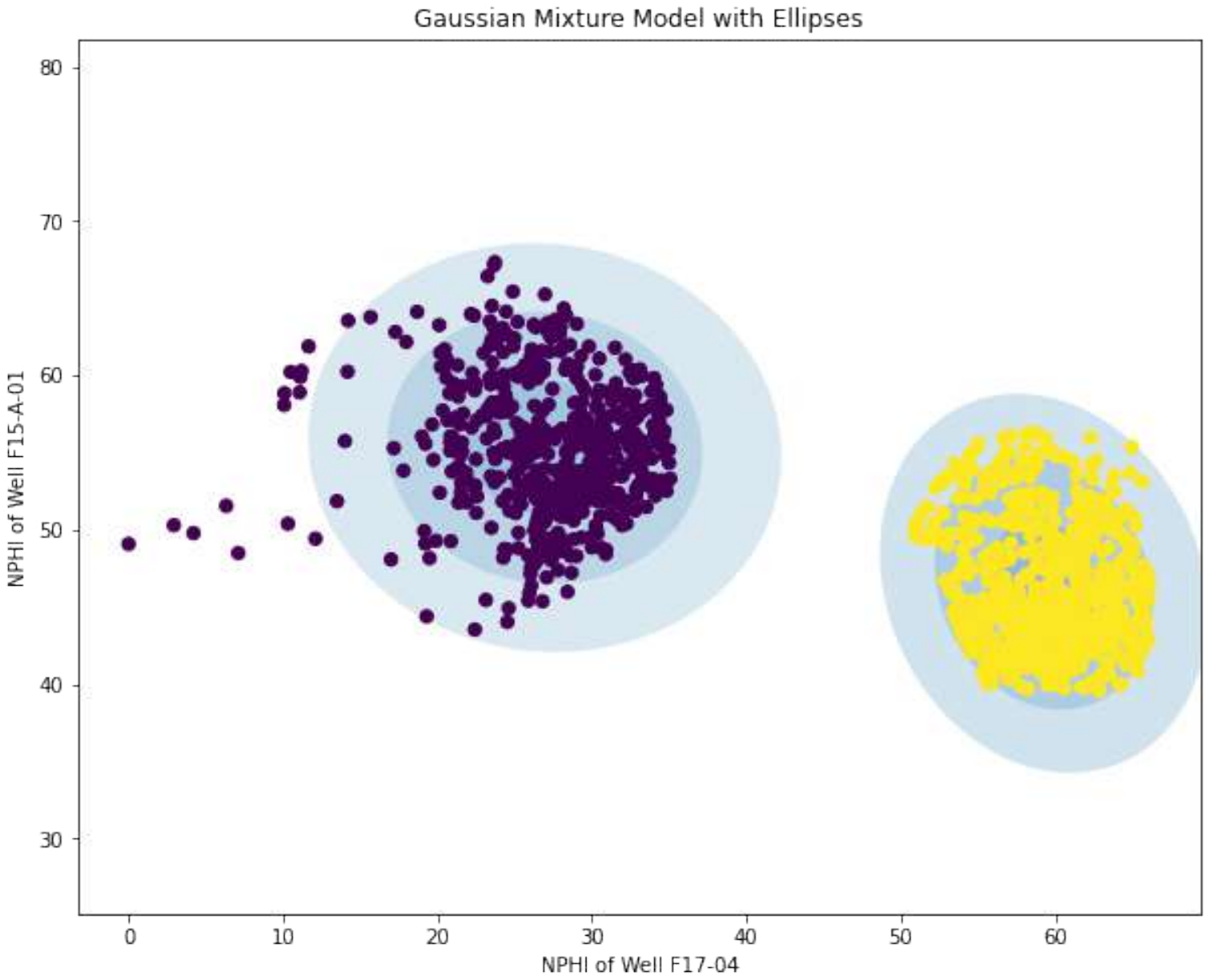}
\end{minipage} &
\begin{minipage}{0.26\textwidth}
    \centering
    \includegraphics[width=\textwidth]{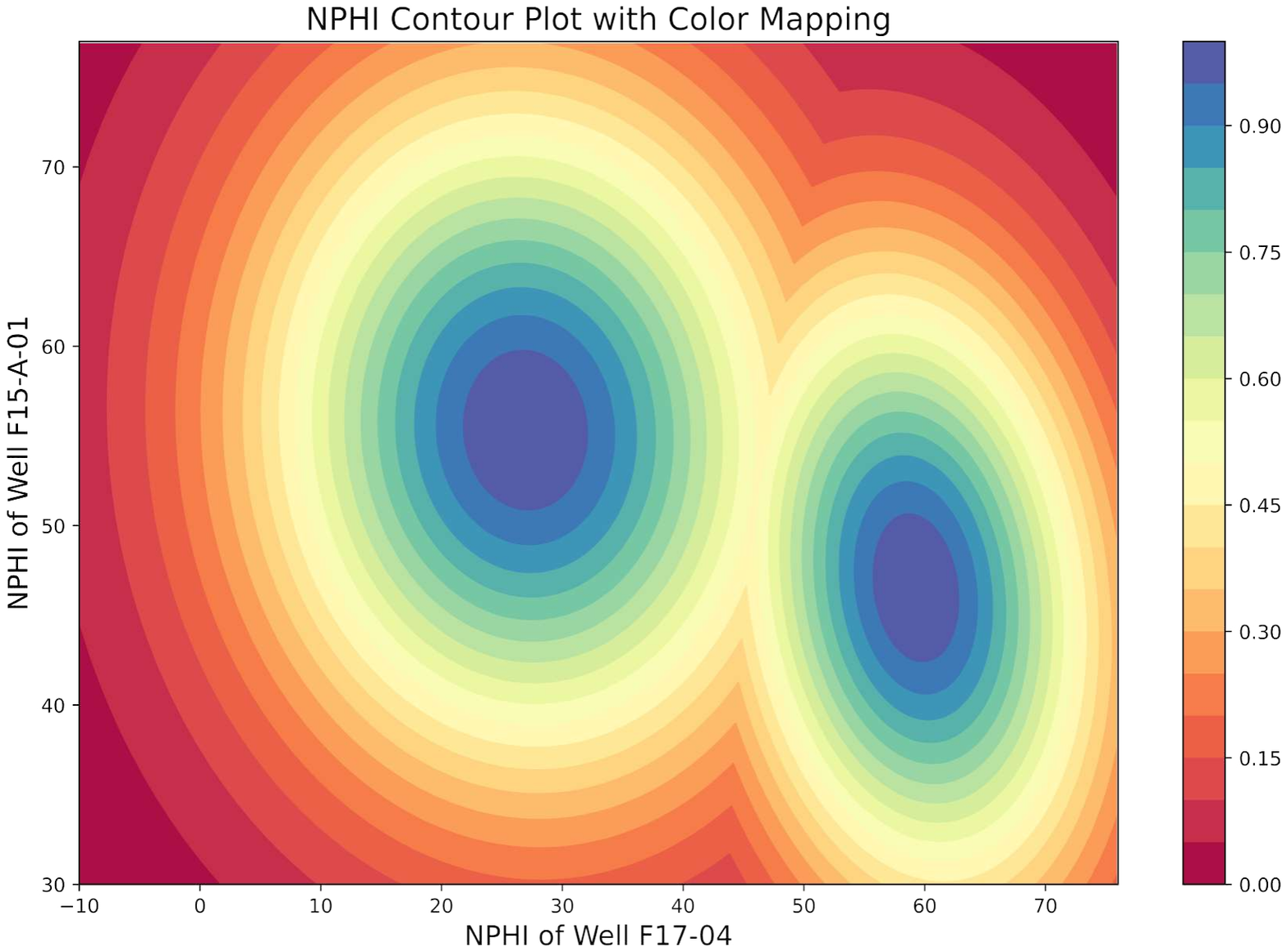}
\end{minipage} &
\begin{minipage}{0.26\textwidth}
    \centering
    \includegraphics[width=\textwidth]{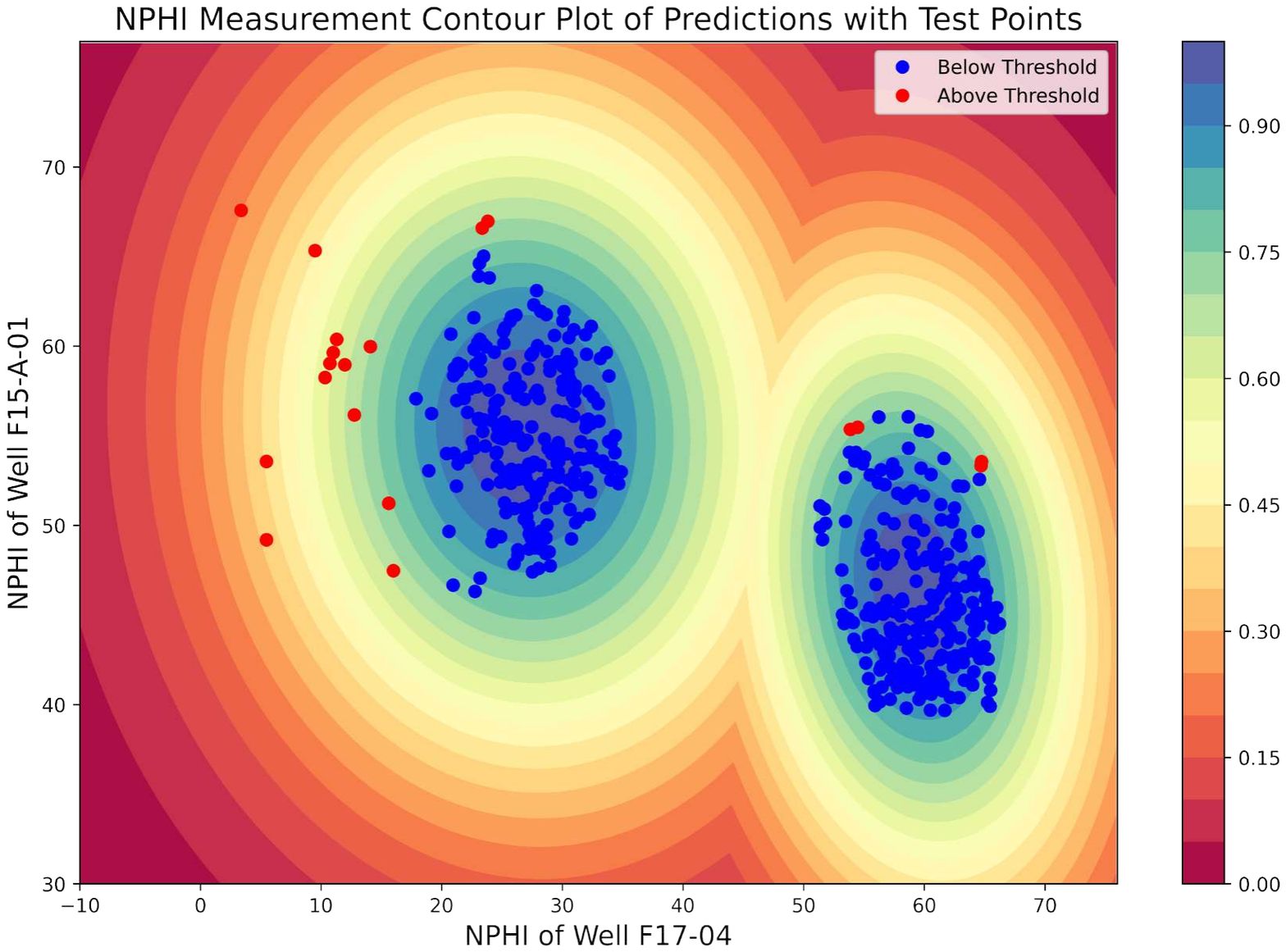}
\end{minipage} \\
\hline
\end{tabular}
\caption{GMM analysis of well log data (GR, DT, RHOB, NPHI). The leftmost columns display scatter plots of the datasets, showing normal and anomalous data points. The second column presents the GMM during the training stage, with ellipses representing the fitted data distribution. The third column depicts the data distribution as interpreted by the GMM, using contour plots with color mapping. The fourth column highlights the testing stage, showing the model’s performance in detecting anomalies, marked in red.}

\end{figure}

The scatter plots in the first column illustrate the anomaly detection results using the IF algorithm, as explained in the methodology section. The IF method effectively identifies anomalies in the GR, DT, RHOB, and NPHI datasets, marking outliers in red. The distribution of normal and anomalous points shows clear clustering, with most anomalies located at the edges of the data distribution. The data distributions identified in this stage will be used for the GMM in subsequent steps, where the model will focus on modeling the structure of the data distributions and detecting anomalies that deviate from these distributions.

The second column illustrates the training stage of the GMM, where the model fits Gaussian ellipses around the data clusters for each dataset (GR, DT, RHOB, NPHI). Since the dataset is unlabeled, the GMM performs unsupervised clustering, grouping the data points based on their inherent distributions. The model fits two Gaussian components to the data, as shown by the yellow and purple ellipses, which represent the high probability regions for each cluster. These ellipses capture the underlying structure of the data, effectively distinguishing between the normal data points in each dataset. The GMM’s training process enables it to model the data distribution, with the ellipses visualizing how the model clusters the data into two groups based on the learned parameters. The resulting model will later be used to identify anomalies by detecting points that deviate from these well-defined data distributions.

The third column presents the contour plots visualizing the probability density functions (PDFs) of the data distributions learned by the GMM. These plots show the regions of higher and lower probability, with contour levels indicating the model's certainty in different areas. The test data points are overlaid on the contour plots to highlight their position relative to the defined probability threshold. Points in higher-density regions are considered normal, while those in lower-density areas are detected as anomalies. The contour plots reveal how well the GMM captures the data structure and how the model separates the clusters, demonstrating its ability to model data distributions effectively. These visualizations also provide insights into the areas where further refinement may be needed, particularly in distinguishing between closely related data points or detecting anomalies that fall outside the primary data distributions.

The fourth column presents the testing stage of the GMM, where the test data is evaluated and visualized alongside the model's predictions. The contour plots show the GMM's predicted probability density, with test data points overlaid. Points are color-coded based on whether they fall above or below a defined probability threshold, indicating their position in high-density (normal) or low-density (anomalous) regions. Points above the threshold are marked in blue (normal), and those below are marked in red (anomalies). These visualizations highlight how well the GMM separates the data into distinct density regions and show the model's ability to distinguish between normal data and anomalies. Comparing results across datasets (GR, DT, RHOB, NPHI), the DT dataset exhibits a clear concentration of anomalies in low-density regions, aligning well with the GMM’s predictions. However, datasets like GR and NPHI exhibit some misclassifications of borderline points, suggesting that further threshold adjustments may be necessary. While the GMM effectively models data distributions, further refinement is required to improve anomaly detection, especially for overlapping data.

\subsection{Ensemble generative adversarial networks analysis}

The EGAN analysis for the GR, RHOB, DT, and NPHI logs, as shown in Figure 4, demonstrates the model’s ability to effectively approximate the underlying data distributions and identify anomalies. 

\begin{figure}[htbp]
\centering
\begin{tabular}{|c|c|c|c|}
\hline
\textbf{Datasets} & \textbf{Training Stage} & \textbf{Data Distribution} & \textbf{Testing Stage} \\
\hline
\begin{minipage}{0.24\textwidth}
    \centering
    \includegraphics[width=\textwidth]{pd_Figures/Figure1_a.pdf}
    {\textbf{GR}}
\end{minipage} &
\begin{minipage}{0.26\textwidth}
    \centering
    \includegraphics[width=\textwidth]{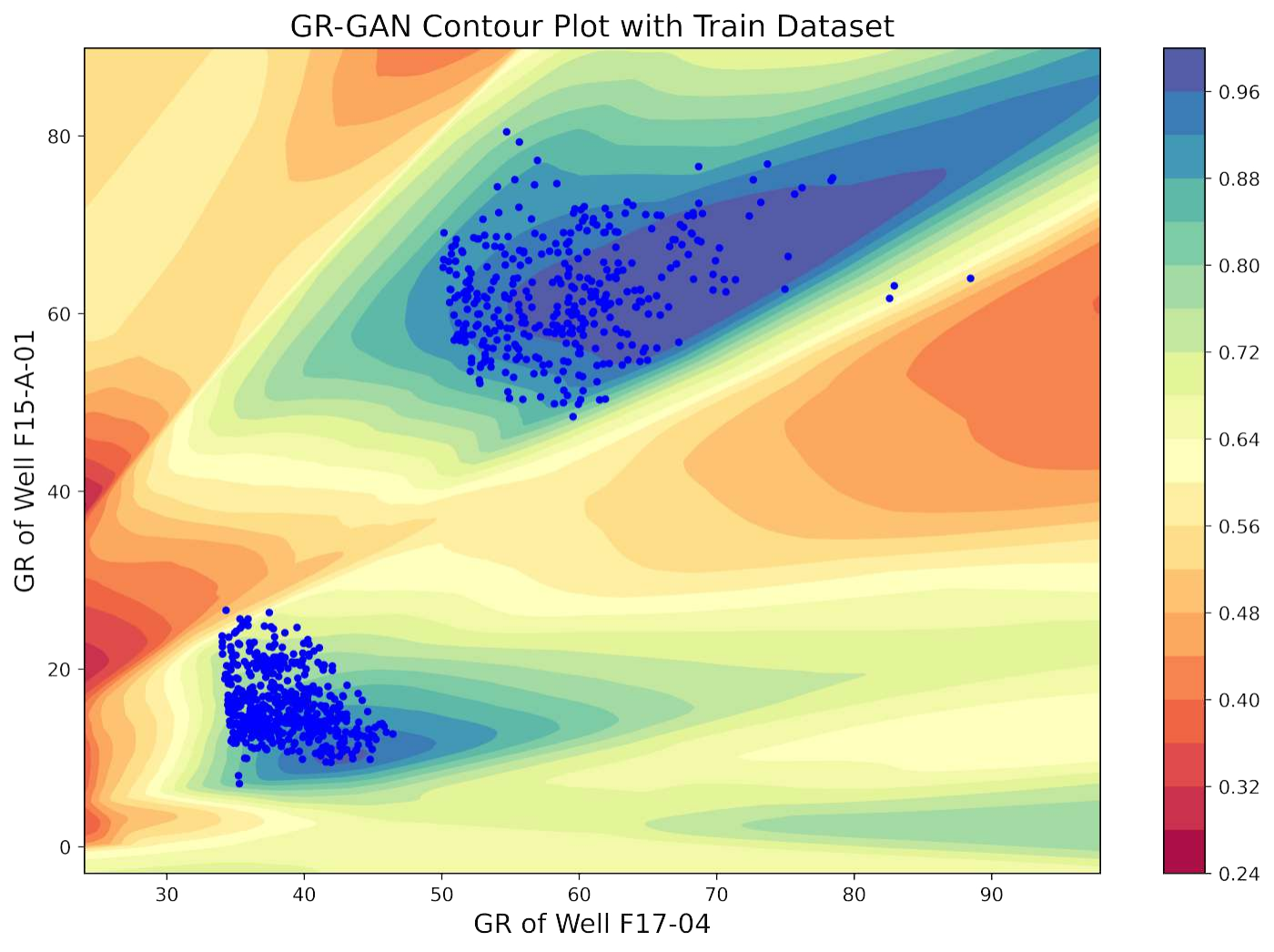}
\end{minipage} &
\begin{minipage}{0.26\textwidth}
    \centering
    \includegraphics[width=\textwidth]{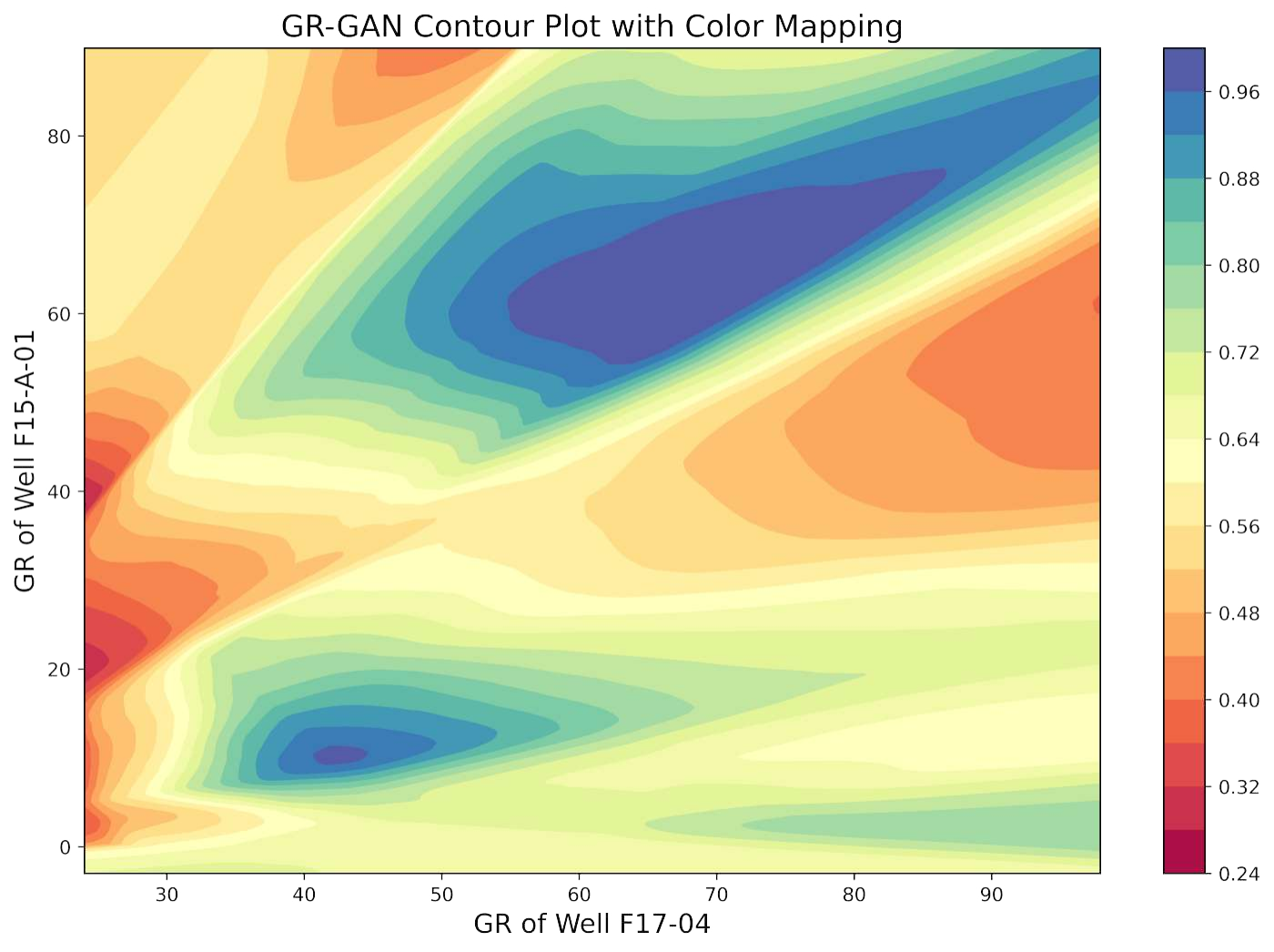}
\end{minipage} &
\begin{minipage}{0.26\textwidth}
    \centering
    \includegraphics[width=\textwidth]{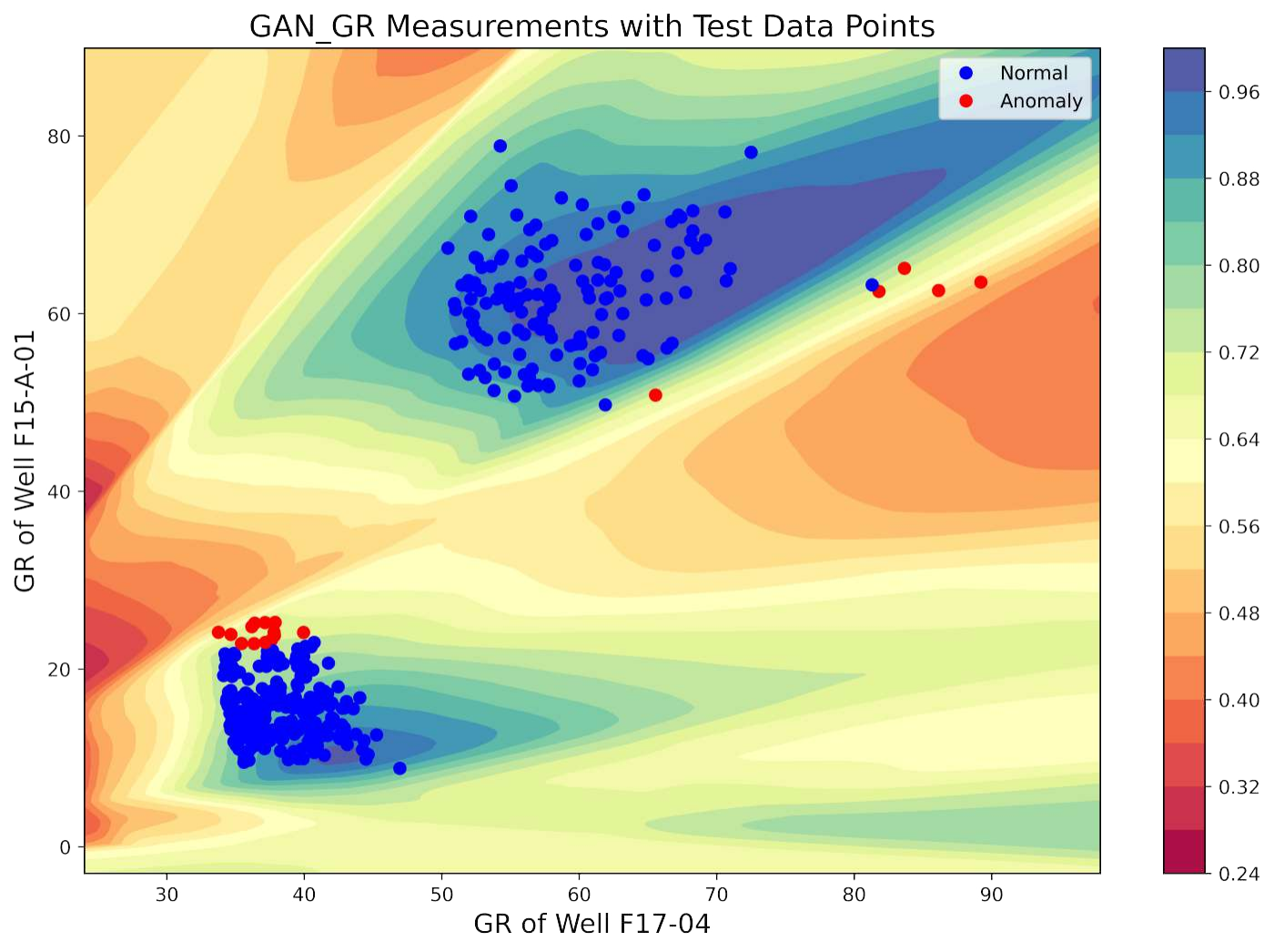}
\end{minipage} \\
\hline
\begin{minipage}{0.24\textwidth}
    \centering
    \includegraphics[width=\textwidth]{pd_Figures/Figure1_c.pdf}
    {\textbf{RHOB}}
\end{minipage} &
\begin{minipage}{0.26\textwidth}
    \centering
    \includegraphics[width=\textwidth]{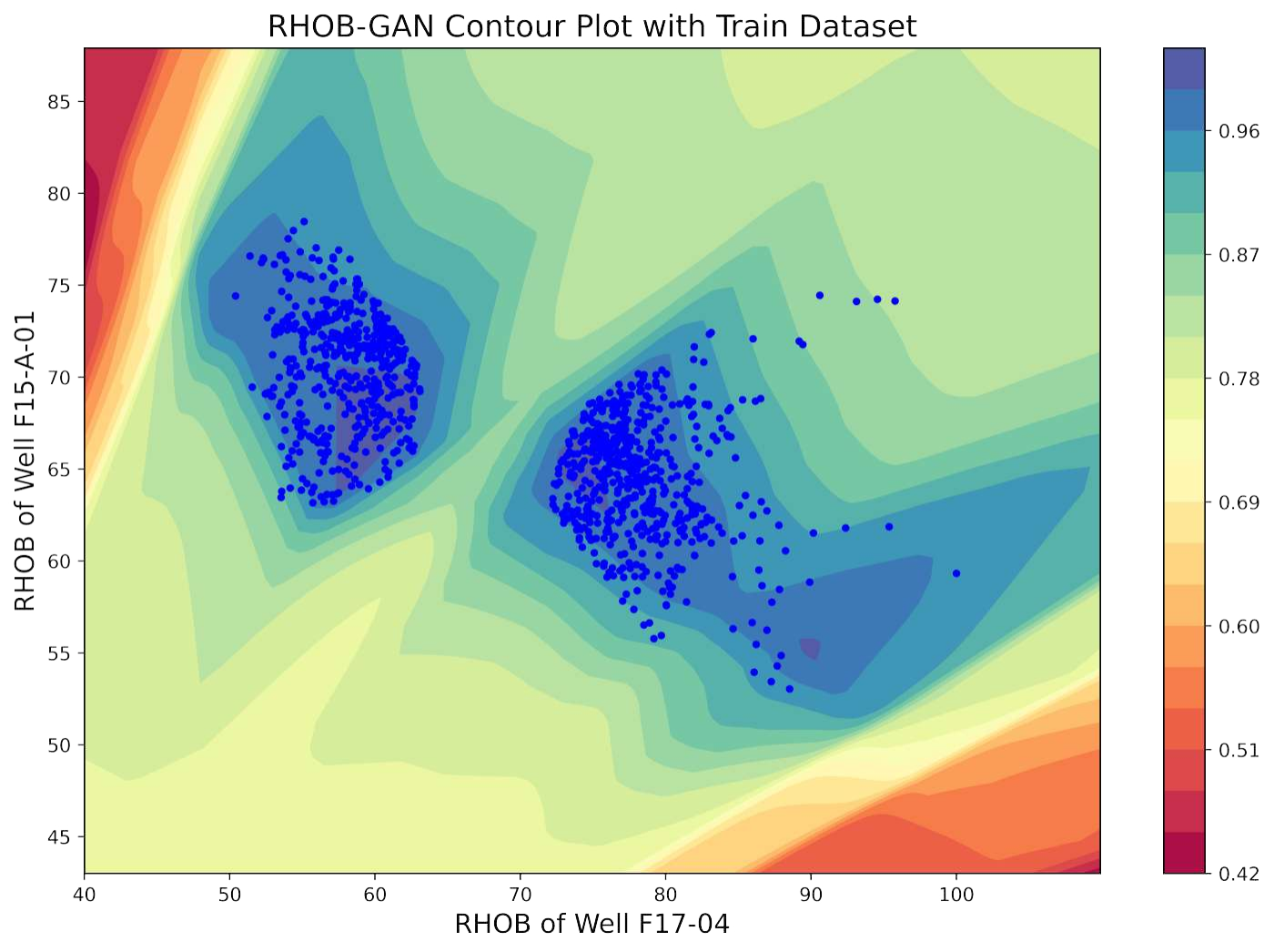}
\end{minipage} &
\begin{minipage}{0.26\textwidth}
    \centering
    \includegraphics[width=\textwidth]{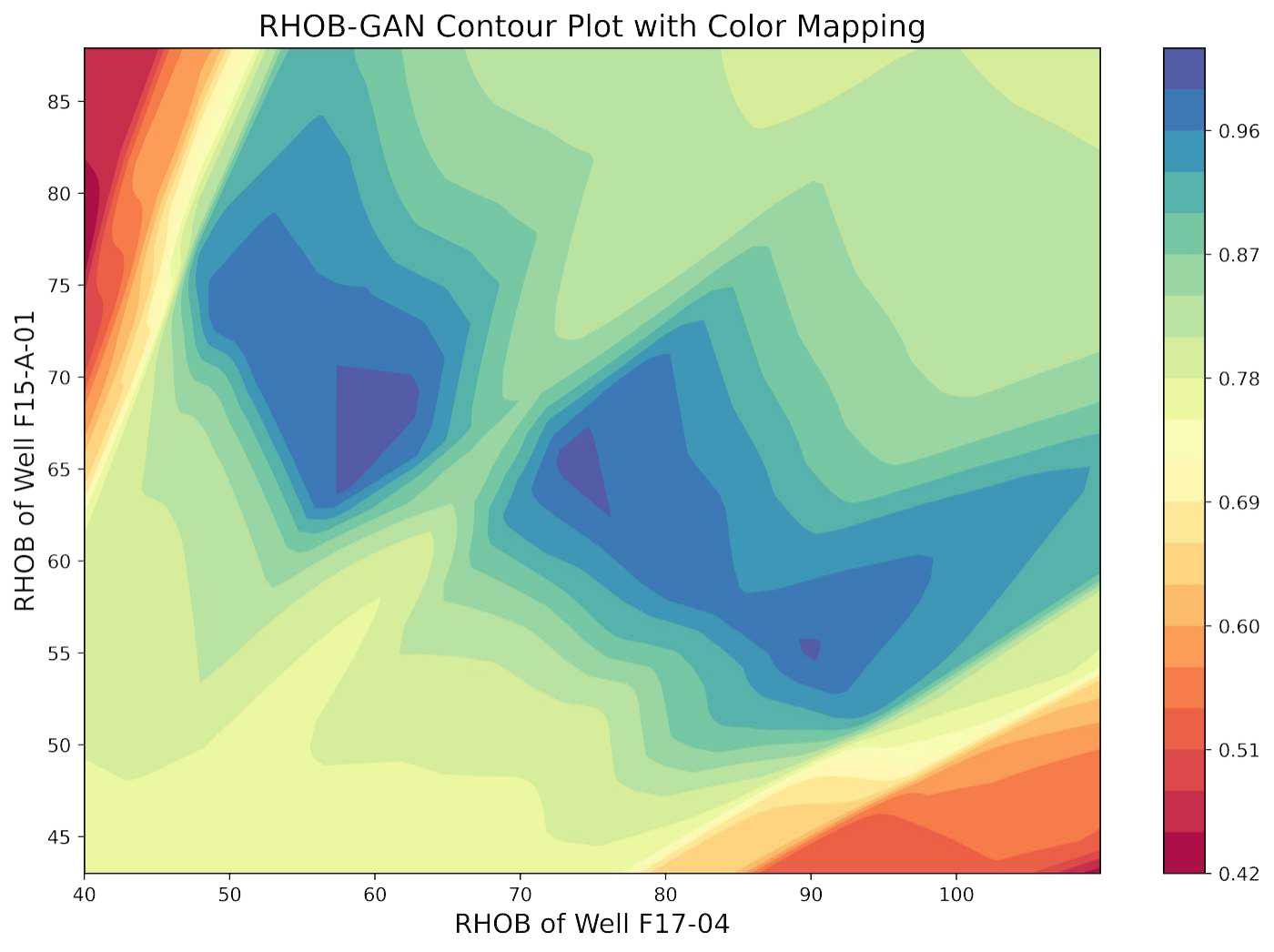}
\end{minipage} &
\begin{minipage}{0.26\textwidth}
    \centering
    \includegraphics[width=\textwidth]{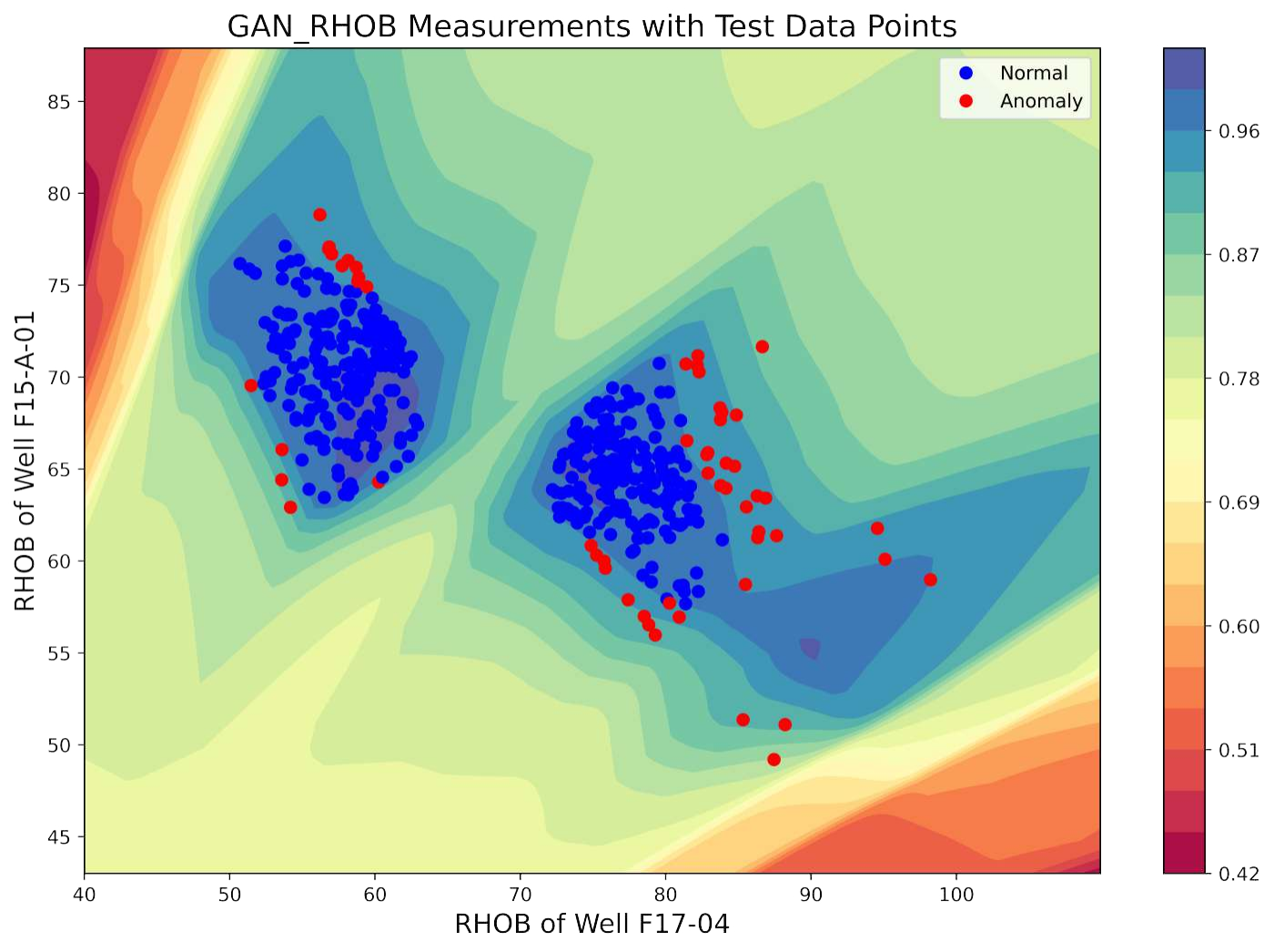}
\end{minipage} \\
\hline
\begin{minipage}{0.24\textwidth}
    \centering
    \includegraphics[width=\textwidth]{pd_Figures/Figure1_b.pdf}
    {\textbf{DT}}
\end{minipage} &
\begin{minipage}{0.26\textwidth}
    \centering
    \includegraphics[width=\textwidth]{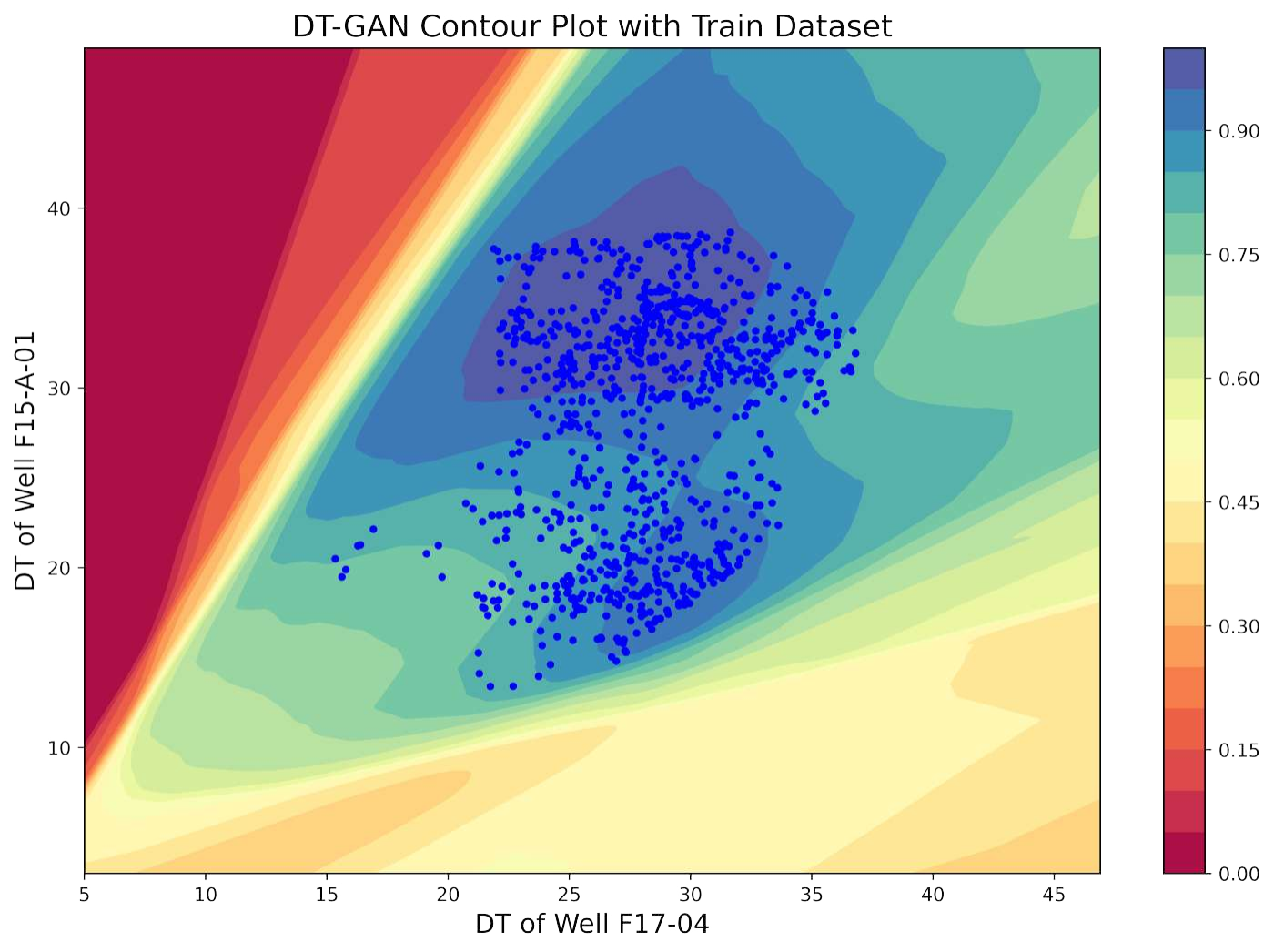}
\end{minipage} &
\begin{minipage}{0.26\textwidth}
    \centering
    \includegraphics[width=\textwidth]{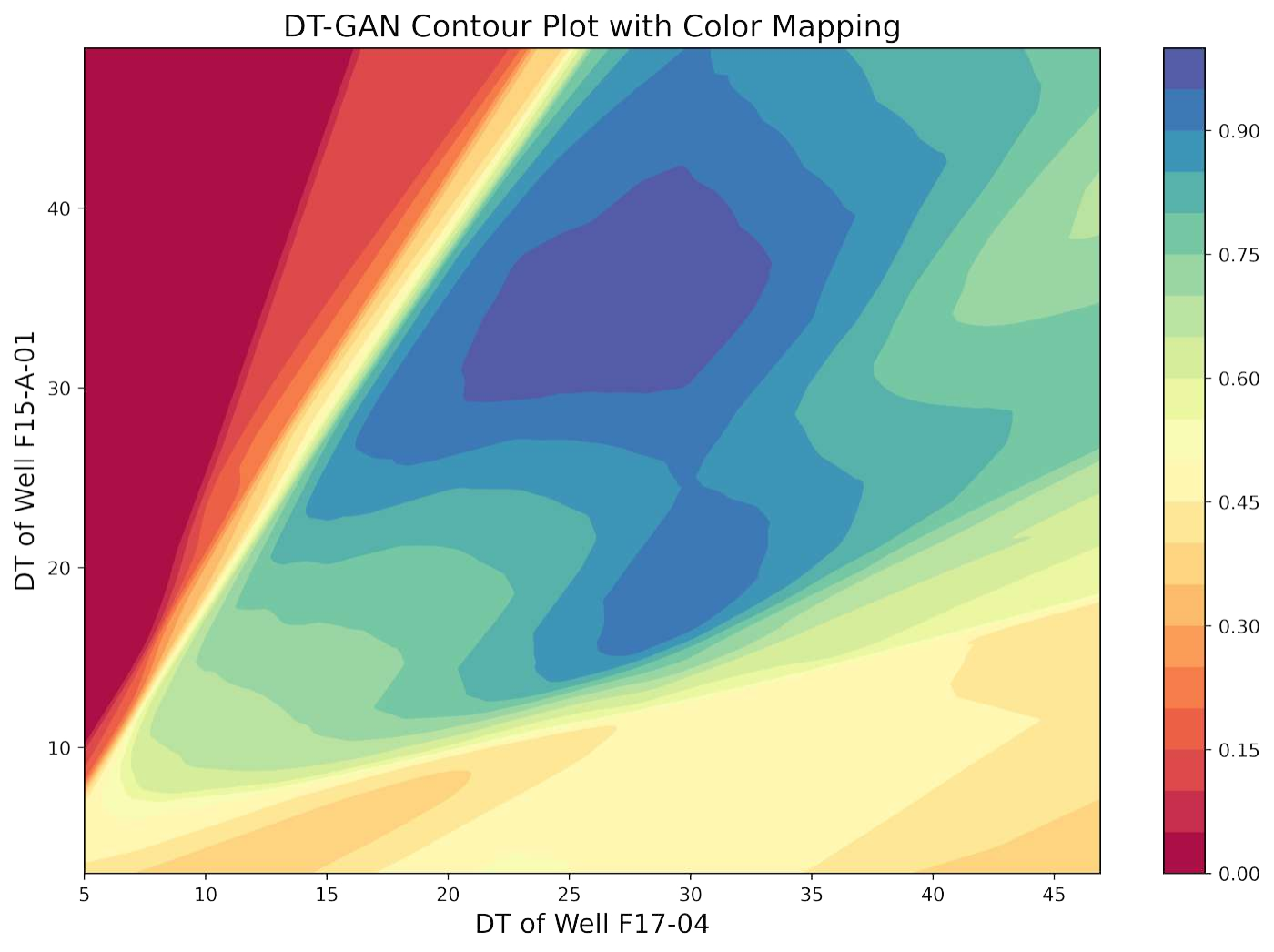}
\end{minipage} &
\begin{minipage}{0.26\textwidth}
    \centering
    \includegraphics[width=\textwidth]{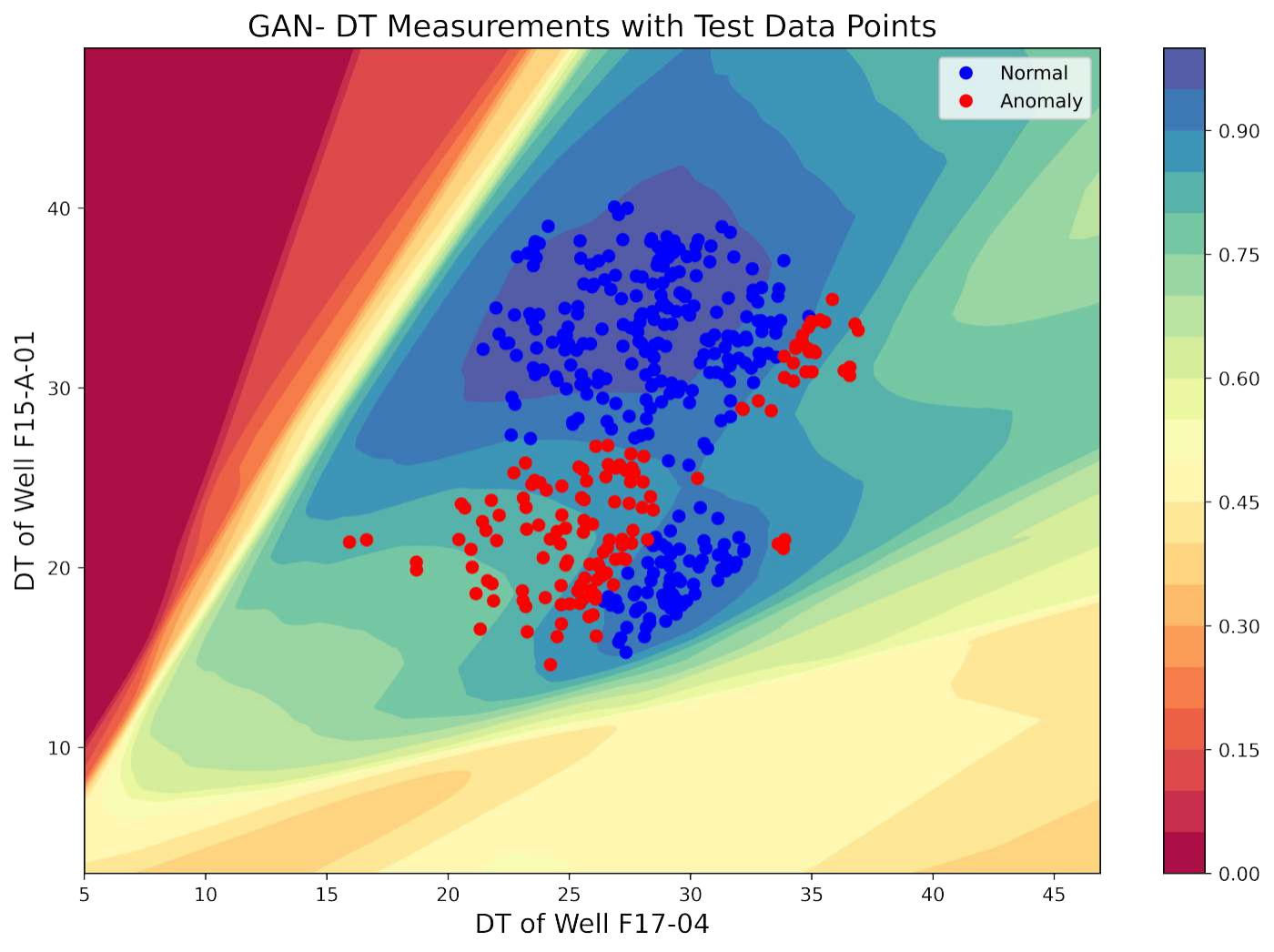}
\end{minipage} \\
\hline
\begin{minipage}{0.24\textwidth}
    \centering
    \includegraphics[width=\textwidth]{pd_Figures/Figure1_d.pdf}
{\textbf{NPHI}}
\end{minipage} &
\begin{minipage}{0.26\textwidth}
    \centering
    \includegraphics[width=\textwidth]{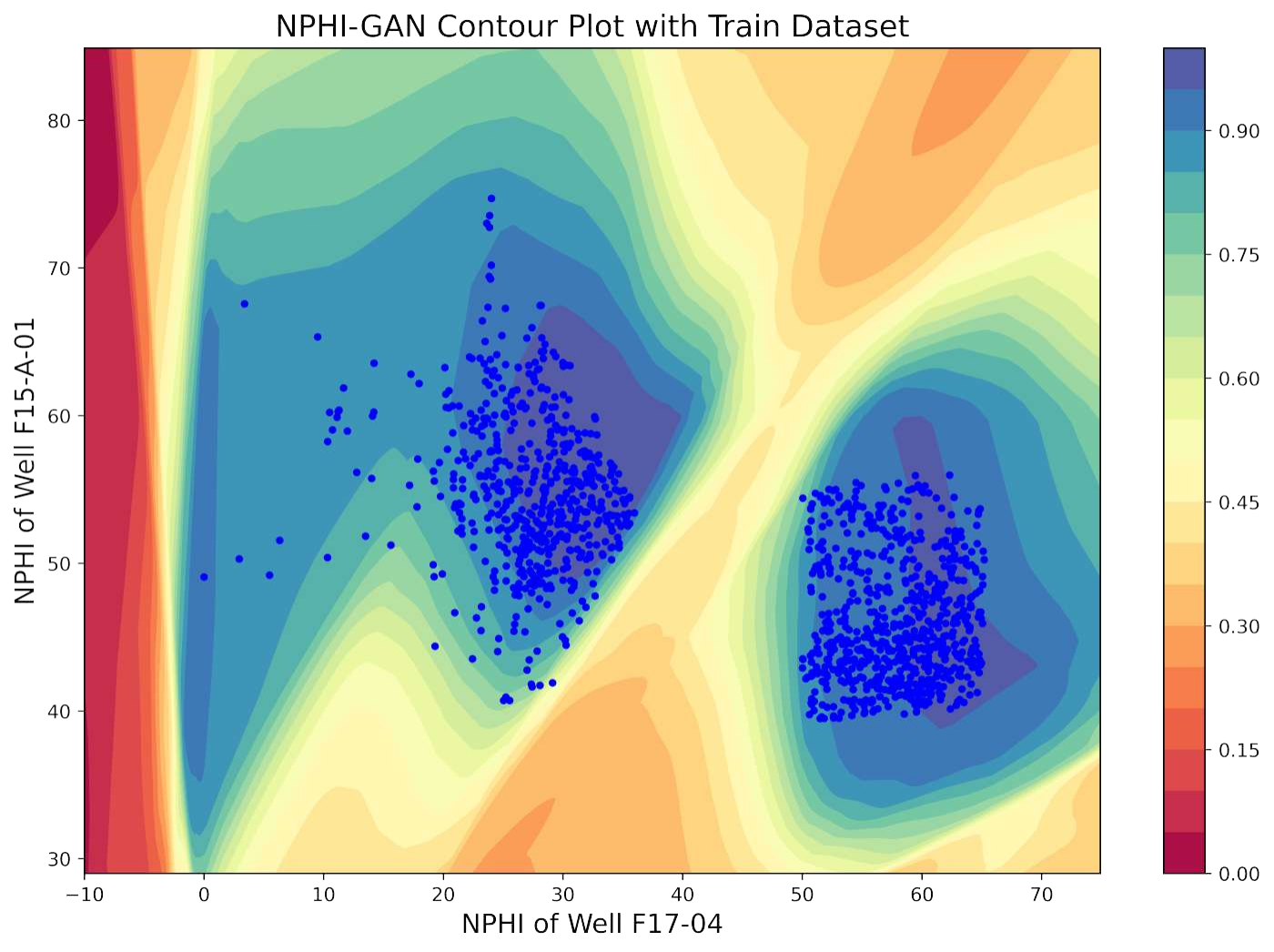}
\end{minipage} &
\begin{minipage}{0.26\textwidth}
    \centering
    \includegraphics[width=\textwidth]{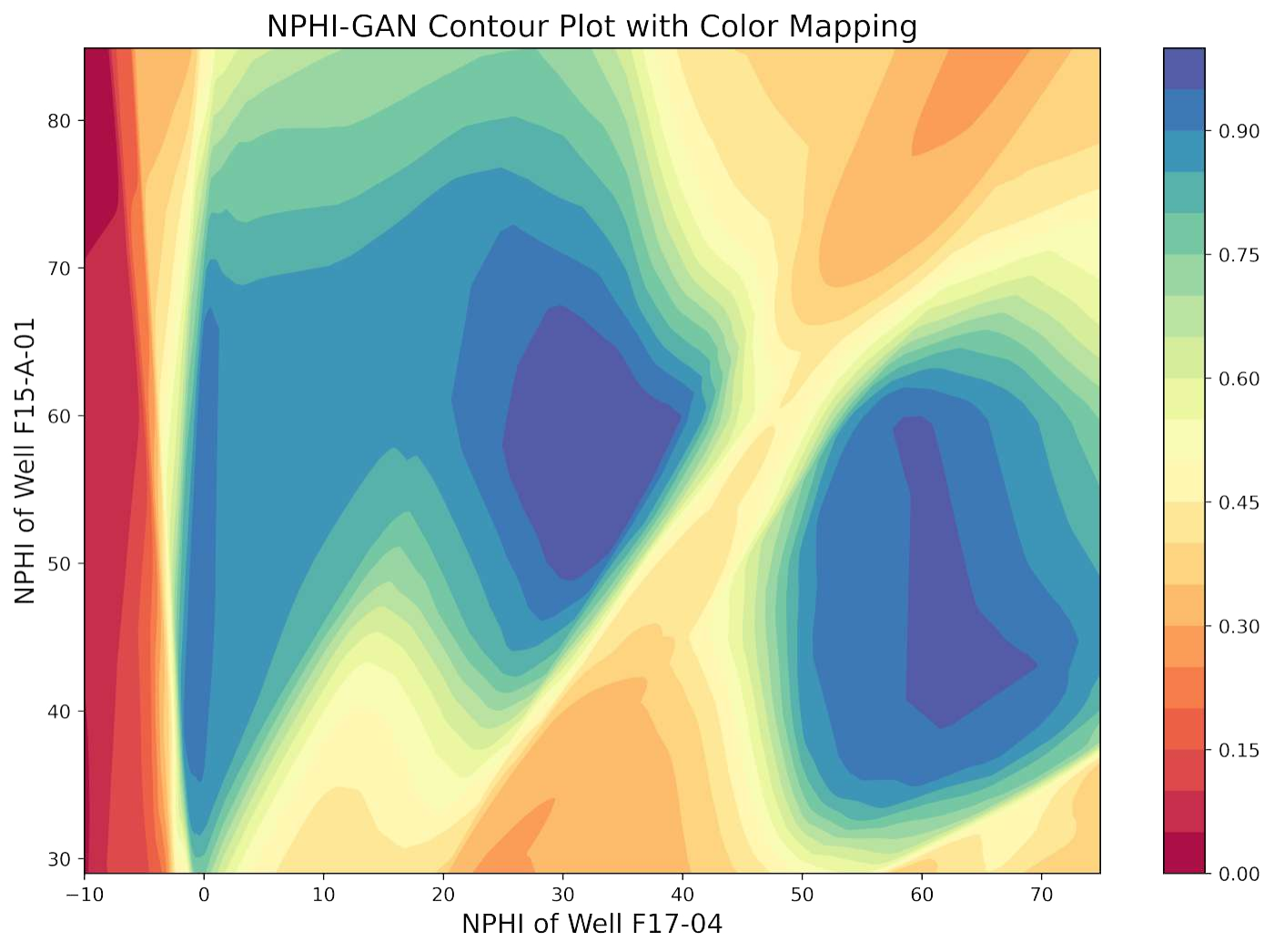}
\end{minipage} &
\begin{minipage}{0.26\textwidth}
    \centering
    \includegraphics[width=\textwidth]{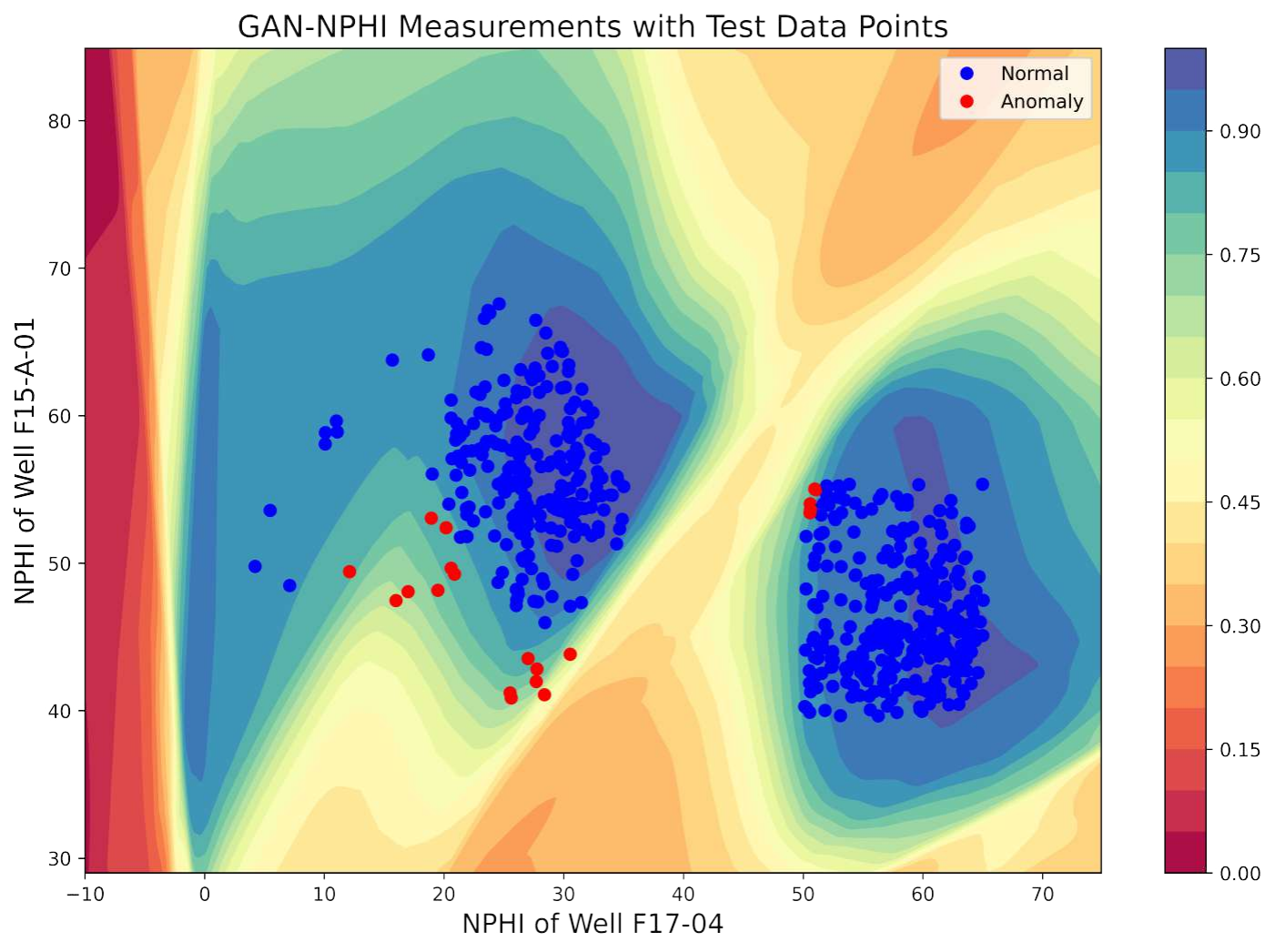}
\end{minipage} \\
\hline
\end{tabular}
\caption{ EGAN analysis of well log data (GR, RHOB, DT, NPHI). The first column shows scatter plots of the datasets with normal (blue) and anomalous (red) points. The second column presents contour plots of the EGAN’s learned data distribution, highlighting high and low-density regions. The third column visualizes the data distribution, showing probability contours for normal and anomalous data. The fourth column illustrates the model’s performance in detecting anomalies, with points above the threshold in blue (normal) and those below in red (anomalies).}

\end{figure}

The first column provides scatter plots illustrating anomaly detection using the IF algorithm. In these plots, normal data points are marked in blue, and anomalies are marked in red. The clustering of points shows a clear distinction between normal and anomalous data, with anomalies generally located at the edges of the data distributions. This step establishes the foundation for later stages by identifying potential outliers, which are critical for the GAN's learning process.

In the training stage (second column), the contour plots illustrate how the GAN learns the data distribution by modeling high and low-density regions. The GAN captures the underlying structure of the data through smooth gradients and clearly defined boundaries between normal and anomalous regions. These plots show how well the GAN approximates the real data distribution, with high-density areas (normal points) shaded in deeper colors and low-density areas (potential anomalies) shaded in lighter colors.

The data distribution (third column) further validates the model's performance by visualizing the PDFs. The contours show the model’s confidence in different regions, with dense areas representing normal data and sparse areas representing potential anomalies. Test data points are overlaid on these contours, with points inside high-density regions classified as normal, while points outside these areas are considered anomalous. This visualization allows for a clear assessment of the model's ability to separate normal and anomalous data.

In the testing stage (fourth column), the model’s performance is assessed by overlaying test data on the predicted contour plots. Test points are color-coded based on whether they fall above or below a defined probability threshold. Points within high-density regions are marked in blue, indicating normal data, while those outside these regions are marked in red, signifying anomalies. This color-coding highlights the GAN's ability to differentiate between normal data and outliers, providing a clear visual understanding of its anomaly detection performance.

When comparing results across datasets (GR, DT, RHOB, NPHI), the DT dataset exhibits the most distinct separation between normal and anomalous points, with a significant concentration of points in the blue (normal) regions. The GR and NPHI datasets show more overlap between normal and anomalous points, suggesting that further tuning or adjustments to the threshold are needed to improve the GAN’s performance. These results indicate that the GAN is effective at modeling well-separated data distributions but may require additional refinement to handle datasets with overlapping or closely packed data points.

\subsection{Performance comparison of EGAN and GMM for anomaly classification using labeled data }

The performance of EGANs and GMM was evaluated across several datasets using key metrics—precision, recall, and F1 score. Table 4 summarizes the results for both models, showing that EGANs generally outperformed GMM in precision and F1 score, demonstrating better ability to minimize false positives. For example, in the GR dataset, EGANs achieved a higher precision (0.62) and F1 score (0.76), while GMM had a lower precision (0.38) despite high recall (0.95). Similarly, EGANs consistently showed better precision and F1 scores in other datasets such as RHOB, DT, and NPHI, where GMM generally produced more false positives.

\begin{table}[h]
\centering
\caption{Summary of performance metrics for various datasets using EGAN and GMM}
\begin{tabular}{|l|c|c|c|c|c|c|}
\hline
\textbf{Dataset} & \multicolumn{3}{c|}{\textbf{EGANs}} & \multicolumn{3}{c|}{\textbf{GMM}} \\ \hline
                  & Precision & Recall & \textbf{F1} & Precision & Recall & F1 \\ \hline
GR                & 0.62     & 0.97   & \textbf{0.76} & 0.38     & 0.95   & 0.54 \\ \hline
RHOB              & 0.52     & 0.93   & \textbf{0.67} & 0.50     & 0.96   & 0.65 \\ \hline
DT                & 0.70     & 0.87   & \textbf{0.79} & 0.56     & 0.98   & 0.71 \\ \hline
NPHI              & 0.53     & 0.94   & \textbf{0.68} & 0.47     & 0.89   & 0.61 \\ \hline
\end{tabular}
\label{tab:performance_metrics}
\end{table}

It is important to clarify that the classification performed in this study is univariate—each variable (GR, DT, RHOB, NPHI) is analyzed independently for anomaly detection. Although the bivariate distributions and contour plots presented in the figures are useful for visualizing the learned data distributions and model behavior, they do not influence the classification process. These visualizations are meant to provide insights into how the models separate the data in two-dimensional space but are not used in the actual anomaly classification, which is based on individual variables. Thus, the classification task remains univariate, with the bivariate plots serving only as a tool for model evaluation and understanding.

For a more detailed exploration of the code, datasets, and result visualizations used in this study, please visit the following GitHub repository: \href{https://github.com/ARhaman/EGANs-vs.GMM}{https://github.com/ARhaman/EGANs-vs.GMM}. This repository contains Python scripts and Jupyter notebooks for model implementation and evaluation, well log data (GR, DT, RHOB, and NPHI) used in this study, and visualizations and results from the anomaly detection analysis.

\section{Future directions}
This study highlights the effectiveness of EGANs in anomaly detection for well log data. However, there are several avenues for future work that could enhance and expand the findings:

\begin{itemize}
    \item \textbf{Multivariate anomaly detection:} While this study focused on univariate anomaly detection, future research should explore multivariate anomaly detection to leverage the relationships between different well log features such as GR, DT, RHOB, and NPHI. This would provide a more comprehensive approach to detecting complex anomalies that depend on the interactions between multiple variables.
    \item \textbf{Enhanced GAN architectures:} To address the challenges of mode collapse and enhance the stability and performance of the models, future work could incorporate Wasserstein GANs (WGANs) or conditional GANs (cGANs). These architectures are well-suited for improving the learning capacity and data generation quality, which can be critical for high-dimensional well log data.
    \item \textbf{Real-time Anomaly Detection:} In the context of reservoir management, there is a need for real-time anomaly detection. Future research should explore the adaptation of the current model to work in real-time systems, continuously updating anomaly detection as new well log data streams in.
    \item \textbf{Integration with Reservoir Simulation Models:} To maximize the impact of anomaly detection, integrating EGANs into reservoir simulation models can offer a more robust solution. This integration would allow for more accurate predictions and early detection of issues in the subsurface, directly influencing drilling and resource extraction decisions.
    \item \textbf{Hyperparameter optimization:} For EGANs using techniques like Bayesian optimization or grid search would further enhance the model’s ability to learn from well log data. Such optimizations could potentially improve both model performance and computational efficiency.
    \item \textbf{Model interpretability:} It is crucial, especially in applied geoscience fields. Future efforts should focus on developing methods to explain the decisions made by GANs and EGANs using techniques like SHAP (SHapley Additive exPlanations) or LIME (Local Interpretable Model-agnostic Explanations). This will help practitioners better understand the model's reasoning for anomaly classification.
    \item \textbf{Exploration of other anomaly detection techniques:} In addition to EGANs and GMMs, it would be valuable to explore other advanced anomaly detection models such as Autoencoders and Isolation Forests. A comparative study with these techniques can provide insights into which methods are most effective for well log data analysis under different conditions.
\end{itemize}

By pursuing these future directions, researchers can enhance the robustness, accuracy, and applicability of anomaly detection models for well log data in reservoir management, ultimately leading to better decision-making and more efficient use of subsurface resources.

\section{Conclusion}
This research demonstrates the robustness of EGANs in analyzing well log data, providing an effective solution for anomaly detection outside the data distribution and forecasting in geophysics. Extensive testing across various logs highlights the superior performance of EGANs, setting a new benchmark for anomaly detection in geosciences. While this study focuses on univariate anomaly detection, future research should extend to multivariate approaches, leveraging relationships between features like GR, DT, RHOB, and NPHI. EGANs consistently outperformed GMM in precision and F1 scores, reducing false positives while maintaining a strong recall balance. In contrast, GMM, although achieving high recall, often suffered from lower precision, leading to more false positives. These findings emphasize the importance of balancing precision and recall in selecting the appropriate model for anomaly detection tasks. A key advantage of EGANs is their adaptability, allowing for modification of confidence intervals to suit various domains. This flexibility enables users to tailor model sensitivity, making EGANs a robust tool for time series analysis in diverse industrial applications. Additionally, EGANs are well-positioned for real-time anomaly detection, a promising area for future exploration in reservoir management. This work’s novelty lies in comparing GMM’s predictive capabilities with EGANs’ advanced anomaly detection, setting a new standard for well log analysis in the oil and gas industry. Future research should explore other advanced anomaly detection techniques, such as autoencoders and isolation forests, to further enhance model robustness. Additionally, hyperparameter optimization using techniques like Bayesian optimization could improve both computational efficiency and anomaly detection performance.

\section*{Author Contributions}

    \textbf{Abdulrahman Al-Fakih:} Formal analysis, Methodology, Software, Writing – original draft, Data preparation, Code creation. \textbf{A. Koeshidayatullah:} Resources, Supervision, Review \& editing.
    \textbf{Tapan Mukerji:} Conceptualization, Review \& editing, Scientific additions.
    \textbf{SanLinn I. Kaka:} Supervision, reviewed \& edited.

\section*{Declaration of Competing Interest and Use of Generative AI}
The authors affirm that they have no known competing financial interests or personal relationships that may have influenced the work presented in this paper. During the preparation of this work, the author(s) used the ChatGPT language model from OpenAI for refining grammar and enhancing text coherence in this article. After using this tool, the author(s) reviewed and edited the content as needed and take(s) full responsibility for the content of the publication.

\section*{Data Availability}
The codes and datasets used in this study are available on GitHub at \url{https://github.com/ARhaman/EGANs-vs.GMM}. The repository includes Python scripts, Jupyter notebooks for model implementation and evaluation, well log data (GR, DT, RHOB, and NPHI), and visualizations and results from the anomaly detection analysis.

\section*{Acknowledgements}
The authors would like to express their gratitude to the College of Petroleum Engineering at KFUPM for their invaluable support in presenting this work at international conferences. Special thanks are extended to the NLOG website and Utrecht University for providing the dataset. Additionally, the authors acknowledge EAGE for the opportunity to present this work at the European Conference on the Mathematics of Geological Reservoirs (ECMOR 24), held in Oslo, Norway.

\section*{Abbreviations}

\begin{itemize}
    \item AI = Artificial Intelligence
    \item API = Application Programming Interface
    \item cGANs = Conditional Generative Adversarial Networks
    \item D = Discriminator
    \item DT = Sonic Travel Time
    \item EM = Expectation-Maximization (the algorithm used in GMM)
    \item EGAN = Ensemble Generative Adversarial Networks
    \item F1 = F1 Score
    \item FP = False Positives
    \item FN = False Negatives
    \item G = Generator
    \item GAN = Generative Adversarial Networks
    \item GMMs = Gaussian Mixture Models
    \item GR = Gamma Ray
    \item ILD = Deep Resistivity
    \item IF = Isolation Forest
    \item K-Means = Number of Clusters
    \item LIME = Local Interpretable Model-Agnostic Explanations
    \item MLM = Machine Learning Models
    \item MLP = Multi-layer Perceptrons
    \item NPHI = Neutron Porosity
    \item NixtlaClient = A client library provided by Nixtla for interacting with their API
    \item Prec = Precision
    \item PDFs = Probability Density Functions
    \item Rec = Recall
    \item RHOB = Bulk Density
    \item SHAP = SHapley Additive Explanations
    \item SVM = Support Vector Machine
    \item TN = True Negatives
    \item TP = True Positives
    \item Tol = Tolerance for Convergence
    \item URL = Uniform Resource Locator
    \item WGANs = Wasserstein GANs
\end{itemize}



\begin{thebibliography}{1}

\bibitem{adke2022application}
G. Adke.
\newblock Application of GAN for Reducing Data Imbalance under Limited Dataset.
\newblock In {\em Proceedings of the International Joint Conference on Computer Vision, Imaging, and Computer Gamma Ray Graphics Theory and Applications}, pages 4, 2022. https://doi.org/10.5220/0010782800003124.

\bibitem{alfakih2023reservoir}
A. Al-Fakih, S. I. Kaka, and A. I. Koeshidayatullah.
\newblock Reservoir Property Prediction in the North Sea Using Machine Learning.
\newblock {\em IEEE Access}, 11, 2023. https://doi.org/10.1109/ACCESS.2023.3336623.

\bibitem{alfakih2024a}
A. Al-Fakih, A. Koeshidayatullah, and S. Kaka.
\newblock AI-Driven Reservoir Management: GANs and GMM for Enhanced Control.
\newblock In {\em ECMOR 2024}, volume 2024, number 1, pages 1--11, 2024. https://doi.org/10.3997/2214-4609.202437004.

\bibitem{alfakih2024b}
A. Al-Fakih, S. I. Kaka, and A. Koeshidayatullah.
\newblock Utilizing GANs for Synthetic Well Logging Data Generation: A Step Towards Revolutionizing Near-Field Exploration.
\newblock In {\em EAGE/AAPG Workshop on New Discoveries in Mature Basins}, volume 2024, number 1, pages 1--5, 2024. https://doi.org/10.3997/2214-4609.202471016.

\bibitem{alfakih2024c}
A. Al-Fakih, A. Koeshidayatullah, and S. Kaka.
\newblock Enhancing Geoscience Analysis: AI-Driven Imputation of Missing Data in Well Logging Using Generative Models.
\newblock In {\em EGU General Assembly 2024}, Vienna, Austria, 14–19 April 2024, EGU24-10627, 2024. https://doi.org/10.5194/egusphere-egu24-10627.

\bibitem{azizyan2013minimax}
M. Azizyan, A. Singh, and L. A. Wasserman.
\newblock Minimax Theory for High-dimensional Gaussian Mixtures with Sparse Mean Separation.
\newblock In {\em Neural Information Processing Systems}, 2013. https://doi.org/10.48550/arXiv.1306.2035.

\bibitem{bhagyashree2020study}
V. Kushwaha and G. C. Nandi.
\newblock Study of Prevention of Mode Collapse in Generative Adversarial Network (GAN).
\newblock In {\em 4th IEEE Conference on Information and Communication Technology, CICT 2020}, 2020. https://doi.org/10.1109/CICT51604.2020.9312049.

\bibitem{bourou2021review}
S. Bourou, A. El Saer, T. H. Velivassaki, A. Voulkidis, and T. Zahariadis.
\newblock A Review of Tabular Data Synthesis Using GANs on an IDS Dataset.
\newblock {\em Information (Switzerland)}, 12(9), 2021. https://doi.org/10.3390/info12090375.

\bibitem{stauffer1999adaptive}
C. Stauffer and W. E. L. Grimson.
\newblock Adaptive Background Mixture Models for Real-Time Tracking.
\newblock In {\em Proceedings. 1999 IEEE Computer Society Conference on Computer Vision and Pattern Recognition (Cat. No PR00149)}, volume 2, pages 246--252, Fort Collins, CO, USA, 1999. doi: 10.1109/CVPR.1999.784637.

\bibitem{darling2005well}
T. Darling.
\newblock Well Logging and Formation Evaluation.
\newblock In {\em Well Logging and Formation Evaluation}, 2005. https://doi.org/10.1016/B978-0-7506-7883-4.X5000-1.

\bibitem{deoliveira2021synthetic}
L. A. B. de Oliveira and C. de C. Carneiro.
\newblock Synthetic Geochemical Well Logs Generation Using Ensemble Machine Learning Techniques for the Brazilian Pre-Salt Reservoirs.
\newblock {\em Journal of Petroleum Science and Engineering}, 196:108080, 2021. https://doi.org/10.1016/j.petrol.2020.108080.

\bibitem{fernandes2024anomaly}
W. Fernandes, K. S. Komati, and K. Assis de Souza Gazolli.
\newblock Anomaly Detection in Oil-Producing Wells: A Comparative Study of One-Class Classifiers in a Multivariate Time Series Dataset.
\newblock {\em Journal of Petroleum Exploration and Production Technology}, 14(1), 2024. https://doi.org/10.1007/s13202-023-01710-6.

\bibitem{gan2018reservoir}
Y. Gan, J. Cao, Y. Lu, Y. He, and H. Wang.
\newblock Reservoir Prediction Based on Stacked Denoising Auto-Encoder for Feature Extraction.
\newblock In {\em SEG Global Meeting Abstracts}, pages 1736--1739, 2018. https://doi.org/10.1190/IGC2018-426.

\bibitem{goodfellow2014generative}
I. J. Goodfellow, J. Pouget-Abadie, M. Mirza, B. Xu, D. Warde-Farley, S. Ozair, A. Courville, and Y. Bengio.
\newblock Generative Adversarial Nets.
\newblock In {\em Proceedings of the 27th International Conference on Neural Information Processing Systems (NIPS'14)}, pages 2672--2680, MIT Press, 2014. https://dl.acm.org/doi/10.5555/2969033.2969125.

\bibitem{han2021gan}
X. Han, X. Chen, and L.-P. Liu.
\newblock GAN Ensemble for Anomaly Detection.
\newblock In {\em Proceedings of the AAAI Conference on Artificial Intelligence}, volume 35, number 5, pages 4090--4097, 2021. https://doi.org/10.1609/aaai.v35i5.16530.

\bibitem{ibrahim2021precision}
M. Ibrahim.
\newblock Precision vs. Recall: Understanding How to Classify with Clarity.
\newblock Retrieved from https://wandb.ai/mostafaibrahim17/ml-articles/reports/Precision-vs-Recall-Understanding-How-to-Classify-with-Clarity--Vmlldzo1MTk1MDY5, 2021.

\bibitem{gokcesu2019sequential}
K. Gokcesu, M. M. Neyshabouri, H. Gokcesu, and S. S. Kozat.
\newblock Sequential Outlier Detection Based on Incremental Decision Trees.
\newblock {\em IEEE Transactions on Signal Processing}, 67(4), 993--1005, 2019. doi: 10.1109/TSP.2018.2887406.

\bibitem{kazemi2021igani}
A. Kazemi and H. Meidani.
\newblock IGANI: Iterative Generative Adversarial Networks for Imputation with Application to Traffic Data.
\newblock {\em IEEE Access}, 9, 2021. https://doi.org/10.1109/ACCESS.2021.3103456.

\bibitem{klu2023fscore}
K. Klu.
\newblock F-Score: What are Accuracy, Precision, Recall, and F1 Score?
\newblock Retrieved from https://klu.ai/f-score-what-are-accuracy-precision-recall-and-f1-score/, 2023.

\bibitem{lai2024application}
J. Lai, Y. Su, L. Xiao, F. Zhao, T. Bai, Y. Li, H. Li, Y. Huang, G. Wang, and Z. Qin.
\newblock Application of Geophysical Well Logs in Solving Geologic Issues: Past, Present and Future Prospect.
\newblock {\em Geoscience Frontiers}, 15(3), 2024. https://doi.org/10.1016/j.gsf.2024.101779.

\bibitem{landauer2023deep}
M. Landauer, S. Onder, F. Skopik, and M. Wurzenberger.
\newblock Deep Learning for Anomaly Detection in Log Data: A Survey.
\newblock {\em Machine Learning with Applications}, 12, 100470, 2023. https://doi.org/10.1016/j.mlwa.2023.100470.

\bibitem{lim2024future}
W. Lim, K. S. C. Yong, B. T. Lau, and C. C. L. Tan.
\newblock Future of Generative Adversarial Networks (GAN) for Anomaly Detection in Network Security: A Review.
\newblock {\em Computers \& Security}, 139, 103733, 2024. https://doi.org/10.1016/j.cose.2024.103733.

\bibitem{luthi2001geological}
S. M. Luthi.
\newblock Geological Well Logs.
\newblock In {\em Geological Well Logs}, Springer Berlin Heidelberg, 2001. https://doi.org/10.1007/978-3-662-04627-2.

\bibitem{liu2019anomaly}
J. Liu, H. Zhu, Y. Liu, H. Wu, Y. Lan, and X. Zhang.
\newblock Anomaly Detection for Time Series Using Temporal Convolutional Networks and Gaussian Mixture Model.
\newblock In {\em Journal of Physics: Conference Series}, volume 1187, number 4, page 042111, IOP Publishing, 2019. https://doi.org/10.1088/1742-6596/1187/4/042111.

\bibitem{ma2019petrophysical}
Y. Z. Ma.
\newblock Petrophysical Data Analytics for Reservoir Characterization.
\newblock In {\em Quantitative Geosciences: Data Analytics, Geostatistics, Reservoir Characterization and Modeling}, 2019. https://doi.org/10.1007/978-3-030-17860-49.

\bibitem{marti2015anomaly}
L. Martí, N. Sanchez-Pi, J. M. Molina, and A. C. B. Garcia.
\newblock Anomaly Detection Based on Sensor Data in Petroleum Industry Applications.
\newblock {\em Sensors}, 15(2):2774--2797, 2015. https://doi.org/10.3390/s150202774.

\bibitem{mishra2022evaluation}
A. Mishra, A. Sharma, and A. K. Patidar.
\newblock Evaluation and Development of a Predictive Model for Geophysical Well Log Data Analysis and Reservoir Characterization: Machine Learning Applications to Lithology Prediction.
\newblock {\em Natural Resources Research}, 31(6), 2022. https://doi.org/10.1007/s11053-022-10121-z.

\bibitem{peer2017automated}
U. Peer and J. G. Dy.
\newblock Automated Target Detection for Geophysical Applications.
\newblock {\em IEEE Transactions on Geoscience and Remote Sensing}, 55(3), 2017. https://doi.org/10.1109/TGRS.2016.2627245.

\bibitem{powers2007evaluation}
D. M. W. Powers.
\newblock Evaluation: From Precision, Recall and F-Factor to ROC, In-formedness, Markedness \& Correlation.
\newblock Tech. Rep. SIE-07-001, School of Informatics and Engineering, Flinders University, Adelaide, Australia, 2007. https://doi.org/10.48550/arXiv.2010.16061.

\bibitem{rebala2019introduction}
G. Rebala, A. Ravi, and S. Churiwala.
\newblock An Introduction to Machine Learning.
\newblock {\em Springer}, 2019. https://doi.org/10.1007/978-3-030-15729-6.

\bibitem{reynolds2015gaussian}
D. Reynolds.
\newblock Gaussian Mixture Models.
\newblock In {\em Encyclopedia of Biometrics}, pages 196--202. Springer, 2015. https://doi.org/10.1007/978-1-4899-7488-4-196.

\bibitem{snorkel2022improving}
Snorkel AI.
\newblock Improving Upon Precision, Recall, and F1 with Gain Metrics.
\newblock Retrieved from https://snorkel.ai/precision-recall-f1-gain-metrics/, 2022.

\bibitem{struminskiy2019well}
K. Struminskiy, A. Klenitskiy, A. Reshytko, D. Egorov, A. Shchepetnov, A. Sabirov, D. Vetrov, A. Semenikhin, O. Osmonalieva, and B. Belozerov.
\newblock Well Log Data Standardization, Imputation and Anomaly Detection Using Hidden Markov Models.
\newblock In {\em 4th EAGE Conference on Petroleum Geostatistics}, 2019. https://doi.org/10.3997/2214-4609.201902208.

\bibitem{valentin2018estimation}
M. B. Valentín, C. R. Bom, A. L. M. Compan, M. D. Correia, C. M. de Jesus, A. L. de Souza, M. P. de Albuquerque, and E. L. Faria.
\newblock Estimation of Permeability and Effective Porosity Logs Using Deep Autoencoders in Borehole Image Logs from the Brazilian Pre-Salt Carbonate.
\newblock {\em Journal of Petroleum Science and Engineering}, 170:315--330, 2018. https://doi.org/10.1016/j.petrol.2018.06.038.

\bibitem{zhao2024novel}
C. Zhao, J. Zhao, W. Wang, C. Yuan, and J. Tang.
\newblock A Novel Hybrid Ensemble Model for Mineral Prospectivity Prediction: A Case Study in the Malipo W-Sn Mineral District, Yunnan Province, China.
\newblock {\em Ore Geology Reviews}, 168:106001, 2024. https://doi.org/10.1016/j.oregeorev.2024.106001.

\bibitem{zhang2023systematic}
Y. Zhang, R. Zhang, and B. Zhao.
\newblock A Systematic Review of Generative Adversarial Imputation Network in Missing Data Imputation.
\newblock {\em Neural Computing and Applications}, 35(27), 2023. https://doi.org/10.1007/s00521-023-08840-2.

\bibitem{zong2018deep}
B. Zong, Q. Song, M. R. Min, W. Cheng, C. Lumezanu, D. Cho, and H. Chen.
\newblock Deep Autoencoding Gaussian Mixture Model for Unsupervised Anomaly Detection.
\newblock In {\em Proceedings of the International Conference on Learning Representations}, 2018. https://openreview.net/forum?id=SJUdkecgx.


\end{thebibliography}

\end{document}